\documentclass{aa}
\usepackage{amsmath,amssymb}
\usepackage{mathtools,graphicx,float}
\usepackage{txfonts}
\usepackage{CJK}
\usepackage{upgreek}
\usepackage{multirow,diagbox,ulem,comment}
\usepackage[colorlinks=true,citecolor=blue,linkcolor=blue,urlcolor=blue,breaklinks]{hyperref}%
\usepackage{orcidlink}
\usepackage{longtable}
\date{May 2024}

\newcommand{\frbpoppy}{\texttt{frbpoppy}\xspace}

\newcommand{\pccm}{\,pc\,cm$^{-3}$\xspace} % Dispersion measure units

\begin{document}
\begin{CJK*}{UTF8}{gbsn}

\title{Birth and evolution of fast radio bursts: \\Strong population-based evidence for a neutron-star origin}
\titlerunning{Birth and evolution of FRBs -- CHIME/FRB Catalog 1 and \frbpoppy.}

\def\orcid#1{\unskip$\orcidlink{#1}$}
\author{
Yuyang~Wang (王宇阳) \inst{1}\fnmsep\thanks{\tt y.wang3@uva.nl, leeuwen@astron.nl} \orcid{0000-0002-3822-0389}\
 \and
Joeri~van Leeuwen \inst{2}\fnmsep$^\star$ \orcid{0000-0001-8503-6958}\ 
}
\authorrunning{Yuyang Wang and Joeri~van Leeuwen}
\institute{Anton Pannekoek Institute for Astronomy, University of Amsterdam, Science Park 904, 1098 XH Amsterdam, The Netherlands 
\and
ASTRON, the Netherlands Institute for Radio Astronomy, Oude Hoogeveensedijk 4, 7991 PD, Dwingeloo, The Netherlands
}
\abstract{While the appeal of their extraordinary radio luminosity to our curiosity is undiminished,
  the nature of fast radio bursts (FRBs) has remained unclear.
The challenge has been due in part to small sample sizes and limited understanding of telescope selection effects.
We here present the first inclusion of the entire set of one-off FRBs from CHIME/FRB Catalog 1
in \frbpoppy.
Where previous work had to curate this data set, and fit for few model parameters,
we have developed full multi-dimensional Markov chain Monte Carlo
(MCMC) capabilities for {\frbpoppy}
-- the comprehensive, open-science FRB population synthesis code --
that allow us to include all one-off CHIME bursts.
Through the combination of these two advances we now 
find the best description of the real, underlying FRB
population, with higher confidence than before.
We show that $4\pm3\times10^{3}$ one-off FRBs go off every second between Earth and $z=1$; and we provide a mock
catalog based on our best model, for straightforward inclusion in other studies.
We investigate CHIME side-lobe detection fractions, and FRB luminosity characteristics,
to show that some bright, local FRBs are still being missed.
We find strong evidence that FRB birth rates evolve with the star formation rate of the Universe,
even with a hint of a short (0.1$-$1\,Gyr) delay time.
The preferred contribution of the hosts to the FRB dispersion agrees
with a progenitor birth location in the host disk. 
This population-based evidence solidly aligns with magnetar-like burst sources, 
and we conclude FRBs are emitted by neutron stars. 
}

\maketitle

\end{CJK*}

\section{Introduction}
\label{Sec: introduction}
Fast radio bursts (FRBs) are millisecond-duration phenomena visible in the radio band. Their origin is extragalactic, implying very  high energies are involved \citep[see, e.g.,][for reviews]{Cordes2019, Petroff2019, Petroff2022}. 
After a first specimen was discovered in 2007 using single-pulse searches of archival Parkes telescope 
data \citep{Lorimer2007}, 
the bursts were not 
proven beyond a doubt as an astrophysical phenomenon \citep[see][for similar human-generated signals]{Burke-Spolaor_2011,2015MNRAS.451.3933P} until four more events 
were found 
in the next decade \citep{Thornton2013}. However, their physical origins are still unclear after a further decade of studies. In general, 
FRBs are now empirically divided into two categories, one-offs and repeaters, depending on whether 
the burst is observed to recur or not. However, whether these two populations are intrinsically or physically different is still an open question. 
Furthermore, the periodic activity in some repeaters and host galaxy localizations for both classes 
are interesting for investigating their nature \citep[see][and references therein]{Petroff2022}, 
while their use for studying the intergalactic medium is also promising \citep[e.g.,][]{2021Natur.596..505P}.

In the early days of FRB discovery, opportunities for more quantitative studies of the FRB population were limited. 
This is changing: since the Canadian Hydrogen Intensity Mapping Experiment (CHIME/FRB; \citealt{CHIMEFRB2019}) 
finds 13 FRBs during its pre-commissioning runs, the ability to detect 
$\mathcal{O}(1)$ to $\mathcal{O}(10)$ FRBs per sky per day 
has made statistical studies of the FRB population feasible.
The more than 500 FRBs contained in CHIME/FRB Catalog 1 \citep{CHIMEFRB2021}
have been the subject of several initial population studies
\citep[e.g.,][]{Chawla2022,Shin2023,Bhattacharyya2023}.

Population synthesis provides a method to study the intrinsic properties of an astronomical class,
by treating the various observational biases encountered, 
which are hard to remove analytically.
Such studies have previously been carried out in a field akin to FRBs, for understanding neutron stars \citep[e.g.,][]{1977ApJ...215..885T,2010A&A...509A...7V,2014MNRAS.439.2893B}.
In FRB detection and interpretation,
coupling between the 
intrinsic characteristics of the sources and of the detecting instrument
introduces similar, possibly strong biases \cite[see, e.g.,][]{2019MNRAS.487.5753C}.
\frbpoppy\footnote{FRB POPulation synthesis in PYthon; \\ \mbox{\url{https://github.com/TRASAL/frbpoppy}}} is an open source python package for population synthesis of FRBs \citep{2019ascl.soft11009G}. 
 It simulates the intrinsic, cosmic, underlying, parent populations and applies simulated surveys to these sources, 
 to obtain the surveyed, detected, observed populations (as further detailed in Sect.~\ref{Sec: modeling}; see 
 \citealt{Gardenier2019} for a full and in-depth discussion of these terms). 
 Then these simulated observed populations are compared with the real observed populations by the actual telescopes. 
 So far, the latest released version is \frbpoppy 2.1, which includes modeling of both one-off and repeater FRBs,
 and options for Monte Carlo (MC) simulations.
 Results from studying the population of repeating FRBs using \frbpoppy are presented in \citet{Gardenier2021a}.
 There, the authors show that a single source population of repeating FRBs, 
 with some minor correlations of behavior with repetition rate, can describe all CHIME/FRB observations.
 Subsequently,
 \defcitealias{Gardenier2021b}{GL21}\citet[][hereafter: \citetalias{Gardenier2021b}]{Gardenier2021b}
 conducted a multidimensional MC simulation,
 and find a population that 
 optimally reproduced the FRBs detected by the four largest surveys at that time
 (Parkes-High Time Resolution Universe (HTRU), CHIME/FRB, Australian Square Kilometre Array Pathfinder (ASKAP)-Incoherent and Westerbork Synthesis Radio Telescope (WSRT)-Apertif; see, e.g., \citealt{2018Natur.562..386S,2023A&A...672A.117V}).
 Such a multi-survey simulation has the benefit that uncertainties in the individual selection biases possibly
 average out; but it comes with the downsides that the overall results are harder to interpret and adjust.
 Since then, a different, complementary approach has become possible: with
 CHIME/FRB Catalog 1 \citep{CHIMEFRB2021}
 the sample size has increased from $\sim$100 FRBs to many hundreds, all subjected to, in principle, the same survey
 biases \citep{Amiri2022}. This allows for a more straightforward population synthesis with {\frbpoppy}.
 In this work, we include the bursts classified as one-off FRB in that catalog in  {\frbpoppy}.

 The paper is organized as follows. 
In Sect.~\ref{Sec: cat_beam_model}
we  introduce the catalog, and characteristics of the FRBs contained therein;
we  also discuss the CHIME/FRB beam model reproduction that is essential for our
simulations and describe the modeling of the intrinsic population. 
In Sect.~\ref{Sec: methods}, we  introduce the methods of our new Markov chain Monte Carlo (MCMC) simulations and in Sect.~\ref{Sec: results} we present the best-fitting
parameters from different models. In Sect.~\ref{Sec: discussion},
we discuss how these compare against other studies,
we outline how our data and models can be used by others,
we present the implications for the FRB source population, and we describe future work. 
We conclude in Sect.~\ref{Sec: conclusion}.

\section{The input: catalog, populations,  telescope model}
\label{Sec: cat_beam_model}
\subsection{The CHIME/FRB Catalog 1}
\begin{figure}
    \centering
	\includegraphics[width=0.9\linewidth]{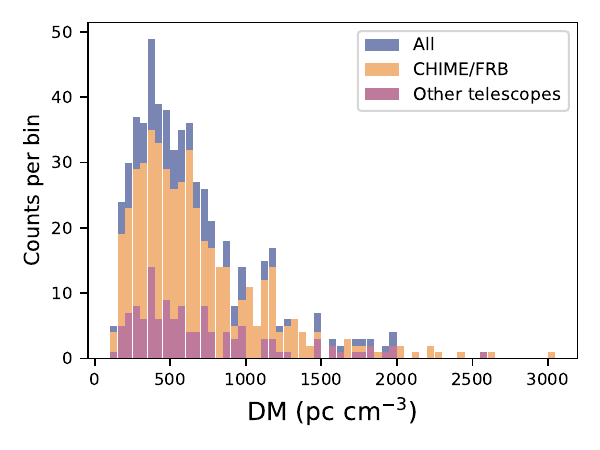}
    \caption{DM distribution histogram of one-off FRBs in the TNS database. The total sample of FRBs is shown in blue while those from CHIME/FRB and other telescopes are shown in orange and green respectively.    }
    \label{Fig: catalog DM}
\end{figure}

\begin{figure}
    \centering
    \includegraphics[width=0.9\linewidth]{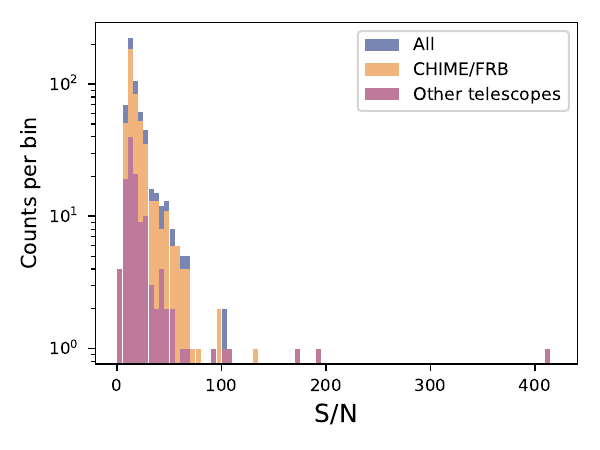}
    \caption{S/N distribution histogram of one-off FRBs, using the same labels and colors as Fig.~\ref{Fig: catalog DM}.}
    \label{Fig: catalog S/N}
\end{figure}

CHIME is a transit radio telescope operating across the 400$-$800\,MHz
band.
It is an excellent
FRB detector, owing to its large field of view, wide bandwidth, and high sensitivity, plus its powerful correlator
\citep{CHIMEFRB2018}. During pre-commissioning, it detected 13 new FRBs \citep{CHIMEFRB2019}. The CHIME/FRB
Catalog 1 was next published in June 2021 \citep{CHIMEFRB2021}, containing 536 FRBs, of which  474 are one-off bursts
and 63 are bursts from 18 repeating sources.
The catalog spans observing dates from 2018 July 25 to 2019 July 1\footnote{In
\frbpoppy the catalog is obtained  from the Transient Name Server (TNS), through a query on these dates.}.
In the current work, we started by limiting ourselves to first reproducing the bursts
that are classified as one-off FRBs.
We stress that \frbpoppy is capable of simulating repeating FRBs too  \citep{Gardenier2021a},
and we will pursue that study in future work.
The catalog is the first set numbering over 100 FRBs
 from a single telescope. That is important because such a set is governed by uniform selection effects.
These selection effects imposed by the differences in 
hardware and in search algorithms \citep{2024Univ...10..158R}
determine, in large part, the overall parameter distributions that describe the observed sample,
 for e.g., the dispersion measures (DMs), distances, bandedness,
and pulse widths.
In Figs.~\ref{Fig: catalog DM} and \ref{Fig: catalog S/N} we compare
two such parameter distributions -- DM and signal-to-noise ratio (S/N) respectively --
for CHIME/FRB 
against those of the pre-existing FRB population discovered by other telescopes.
We here do this as a first impression, to outline the biases at stake,
and investigate these  distributions in more detail later.
The DM distributions are relatively similar.
That is interesting because CHIME/FRB operates at a lower frequency than most other surveys, 
which would generally make higher-DM sources especially hard to find. 
Especially at these low frequencies, such sources
suffer more from intra-channel dispersion and
scatter broadening than low-DM FRBs. 
This disadvantage is possibly mitigated by other relative advantages of the CHIME/FRB system
(its narrow channels might, for example, sufficiently limit the intra-channel dispersion).
It is also possible that other biases, working against low-DM sources instead,
 balance out the selection effects with DM.
Hints for the latter are reported by \citet{Merryfield2023}, who find that 
the CHIME/FRB pipeline selects against bright, low-DM FRBs.
We investigate this in Sect.~\ref{Sec: Side-lobe}. 

For the S/N, the CHIME/FRB distribution is shifted toward higher values than that of the other
telescopes. This could be intrinsic, a result of stricter confidence level requirements, or a combination thereof.

\subsection{Modeling underlying populations}
\label{Sec: modeling}

As the number of detected FRBs increases, a number of sources have shown complex behavior or circumstances. Our goal is
to distill the most important trends from the complicated real FRB population.
We thus used a number of parameters to
model its main characteristics, both source-related properties 
(luminosity, pulse width) as well as population-related properties (number density, DM). Here, we describe the parameter
distribution models adopted in our simulations (see \citealt{Gardenier2019} for further details on these methods).
For reference, the parameters and their meaning are also listed in Table~\ref{Tab:acros}.

\subsubsection{Number density following a power-law model}
\label{sec:density:pl}
In {\frbpoppy}, a number of different models can
establish the number-redshift relation $\mathrm{d} N/\mathrm{d} z$
of intrinsic FRB populations per comoving volume.
In this study, we focused on two models. The first
tracked the cosmic star formation rate (SFR), as we  discuss in the next subsection.
In the second, we created a power-law model that does not depend on SFR. 
In this case,  \frbpoppy modifies the uniform sampling $U(0, 1)$ with exponent $B$
\begin{equation}
    V_{\mathrm{co}, \mathrm{FRB}} = V_{\mathrm{co}, \max } \cdot U(0,1)^{B},
\label{eq:Vco}
\end{equation}
where $B$ determines the slope of the cumulative source-count distribution for detected FRBs, above a certain peak flux density detection threshold, at the high flux density end. 
The parameter $\alpha$ is introduced via
\begin{equation}
    \alpha=-\frac{3}{2 B}.
    \label{eq:alpha}
\end{equation}
In a Euclidean universe, where the FRB count per comoving volume element does not evolve with redshift,
we have $B=1$ and $\alpha=-3/2$.
By introducing  $\alpha$  via this power-law index, it is the same as the $\log N$$-$$\log S$ slope
in the high fluence limit.
From now, we use the term Euclidean model to refer to such non-evolving models with $\alpha=-3/2$.
As a number of studies have suggested or reported deviations from  $\alpha=-3/2$
  (e.g., $\alpha$ = $-$2.2 in both \citealt{James2019} %for the CRAFT sample,
  and \citetalias{Gardenier2021b})
  we do however search for the best-fit value of $\alpha$ in a number of experiments reported in Sect.~\ref{sec:results:pl}.

\subsubsection{Number density following a star formation rate model}
There is ongoing discussion on whether the FRB event rate tracks the SFR.
\citet[][]{James2022b} find it does.
 \cite{Zhang2021} and \cite{Zhang2022}, however, claim that the 
FRB population as a whole does not track the SFR; 
while a delayed SFR model, potentially caused by compact binary mergers, 
cannot be
rejected with the ASKAP and Parkes sample. So, in this work, we considered both the SFR itself,
plus delayed SFR models with three different delay times: 0.1, 0.5, and 1\,Gyr. 
We note that for delay time distribution (DTD), we used a constant model, which assumes all compact binaries have the same delay time $\tau$
(while \citealt{Zhang2021} mention various other DTDs, including power-law, Gaussian or log-normal).
We discuss the results in Sect.~\ref{Sec: SFR} and Sect.~\ref{Sec: res:delayedSFR}.

\subsubsection{Spectral index}
\label{sec:si}
The spectral index $\gamma$
(Eq.~10 in \citealt{Gardenier2019}; called $si$ in  \citetalias{Gardenier2021b})
describes the relative flux density at different frequencies within the spectrum.
In \frbpoppy, this index acts through
converting the FRB fluence into the observing bands of the modeled surveys
(see Eq.~\ref{Eqn: S_peak} describing $S_\text{peak}$). 
It 
is arguably best constrained by comparing FRB fluences and rates from survey at different frequencies.
Such multi-survey simulations are one of the strong points of \frbpoppy
and we will explore this aspect in future work. 
In the current work, however, we only modeled CHIME/FRB, effectively at a single frequency band. 
In this case the resulting $\gamma$ is greatly influenced by the choice of $v_{\text{low}}$, $v_{\text{high}}$, and the
assumption that the power-law relationship ($E_\nu\propto \nu^\gamma$) holds throughout the whole frequency
range. As this degeneracy cannot be solved in a single-frequency study,
we did not treat $\gamma$ as a free parameter here while we formally defined it as $\gamma$ =  $-$1.5. This value is motivated by the mean spectral index found by \citet{Macquart2019} using a sample of 23 FRBs detected with ASKAP.

\subsubsection{Luminosity}
\label{sec:lum}
The bolometric luminosity $L_\text{bol}$ was generated from a power-law distribution with
power-law index $li$
and range $[L_\text{min}, L_\text{max}]$.
Such a  range limit is the simplest method to bound the allowed
  luminosity values and keep the function convergent. Physically,
  it suggests there is some minimum required energy to produce an FRB
  (similar minimum physical luminosities have been proposed for pulsars,
  see e.g.,~\citealt{2010A&A...509A...7V} and references therein)
  and some maximum  available energy
  reservoir per burst.
Since the publication of \citetalias{Gardenier2021b}, the detection of FRB\,20220610A has raised
the maximum observed rest-frame burst
luminosity to  $10^{46}$\,erg\,s$^{-1}$ \citep{2022arXiv221004680R},
so we have increased  $L_\text{max}$ to the same value.
Inferred luminosities depend on the assumed bandwidth of emission;
so some caution is required when comparing these numbers.
To provide context to this extreme luminosity, we note that
the magnetic Eddington limit of a strong-field  ($>3\times 10^{14}$\,G) neutron star
is $10^{42}$\,erg\,s$^{-1}$  \citep{Thompson1995}.
Whether a lower limit to the one-off FRB luminosity exists is unclear.
The $\sim$10$^{37}$\,erg\,s$^{-1}$ luminosity of nearby repeating FRB\,20200120E
\citep{2021ApJ...910L..18B}
already indicates the underlying process, if the same, can operate over a range of brightness.
To limit computing time our simulations require a lower bound though, and for the one-off FRBs under consideration here,
we chose $L_\text{min} = 10^{41}$\,erg\,s$^{-1}$.
This is 3 orders of magnitude below the least luminous localized one-off bursts,
such as FRB\,190608 \citep{Macquart2020} at
$\sim$$10^{44}\,$erg s$^{-1}$
(although given the current state-of-the art,
only relatively bright FRBs can generally be localized).
As it is also conservative compared to the $L_\text{min}=7\times 10^{42}$\,erg\,s$^{-1}$ suggested by
\cite{Cui2022} for their CHIME/FRB non-repeater sample,
we are confident our simulation coverage is  complete.

The luminosity we model in \frbpoppy represents the isotropic bolometric radio luminosity.
  In contrast to similar studies for pulsars \citep[e.g.,][]{2004IAUS..218...41V}
  we did not (yet) include beaming effects for FRBs,
  given the uncertainly of the progenitor models in this regard.
  We discuss this in more detail in Sect.~\ref{Sec:nProgn}.

Other luminosity functions exist,
and could potentially be implemented in the future. Once the number
of detected FRBs allows for it, one could then distinguish between multiple luminosity models.
\citet{Luo2020}, for example, consider a 
Schechter luminosity function.
We do note here that the original strength of this function -- 
for galaxies --
was that it
was derived from self-similar galaxy formation models, 
and next well described the observed galaxy distributions \citep{Schechter1976}.
Its current application to FRBs, in contrast, is ad hoc,
and similar to application of a generic broken power law.
Both functions still require an $L_\text{min}$ lest they diverge at the
low-luminosity end.
Testing a log-normal distribution might also be interesting.
Considerations on the luminosity functions in \frbpoppy are discussed in more depth in
%e.g., Sect.~3.2 of
\citet{Gardenier2019}.
  
\subsubsection{Pulse width}
\label{sec:wint}
The intrinsic pulse width $w_\text{int}$ is modeled with a log-normal distribution
\begin{equation}
    p(w_\text{int})=\frac{1}{w_\text{int}} \cdot \frac{1}{\sigma \sqrt{2 \uppi}} \exp \left(-\frac{(\ln w_\text{int} - \mu)^{2}}{2 \sigma^2}\right),
\label{eq:wint}
\end{equation}
where $\mu,\,\sigma$ are related to the input parameters $w_\text{int, mean}$ and $w_{\text{int, std}}$ by $w_\text{int, mean} = \exp\left(\mu+\sigma^2/2\right)$ and $w_\text{int, mean}=\sqrt{[\exp(\sigma^2)-1] \exp(2\mu+\sigma^2)}$
.

The pulse width $w_\text{arr}$ of an FRB arriving at Earth is
\begin{equation}
    w_\text{arr} = (1 + z)\,w_\text{int}.
    \label{eq:warr}
\end{equation}

The observed pulse width is 
\begin{equation}
w_{\mathrm{eff}}=\sqrt{w_{\mathrm{arr}}^{2}+t_{\mathrm{scat}}^{2}+t_{\mathrm{DM}}^{2}+t_{\mathrm{samp}}^{2}},
\label{eq:weff}
\end{equation}
where $t_{\mathrm{scat}}$ is the scattering time while $t_{\mathrm{samp}}$ and $t_{\mathrm{DM}}$ are the result of instrumental broadening. In our simulation, a log-normal distribution for $t_{\mathrm{scat}}$ is adopted (see Fig.~\ref{Fig: pulse width}, right panel);  $t_{\mathrm{samp}}$ is 0.98304\,ms as in CHIME/FRB Catalog 1; and
$t_{\mathrm{DM}}$ is the intra-channel dispersion smearing the CHIME/FRB back-end incurs for each FRB.

\subsubsection{Dispersion measure}
The total DM of an FRB is 
\begin{equation}
    \mathrm{DM}_{\mathrm{total}} = \frac{\mathrm{DM}_{\text {host}}}{1 + z} + \mathrm{DM}_{\mathrm{IGM}} + \mathrm{DM}_{\mathrm{MW}},
    \label{eq:DMs}
\end{equation}
where $\mathrm{DM}_{\text {host}}$ is the host galaxy contribution (including the circumburst environment contribution
$\mathrm{DM}_{\text {src}}$ and Milky Way halo contribution $\mathrm{DM}_{\text {MW,halo}}$) in the source rest
frame.
We generated $\mathrm{DM}_{\text {host}}$ using a log-normal distribution  ($\ln\mathrm{DM}_{\text {host}} \sim
N(\sigma, \mu)$),
where input parameters $\mathrm{DM}_{\text {host, mean}}$ and $\mathrm{DM}_{\text {host, std}}$ represent the mean and standard deviation of the log-normal distribution respectively; $\mathrm{DM}_{\text {MW}}$ is the Milky-Way interstellar medium (ISM) contribution obtained from the NE2001 model \citep{Cordes2002}. 

The intergalactic medium (IGM) contribution $\mathrm{DM}_{\mathrm{IGM}}$ is estimated using a linear relation with redshift
$\mathrm{DM}_{\mathrm{IGM}} \simeq \mathrm{DM}_{\mathrm{IGM}, \text {slope}}\,z$ \citep{Zhang2018,Macquart2020},
around 
which we 
include the spread from  
 a normal distribution, leading to
\begin{equation}
    \mathrm{DM}_{\mathrm{IGM}} = N\left(\mathrm{DM}_{\mathrm{IGM}, \text {slope}}\,z, 0.2\,\mathrm{DM}_{\mathrm{IGM}, \text {slope}}\,z\right).
\end{equation}

As the value of DM$_{\mathrm{IGM}, \text {slope}}$ derived  in  \citet[][their Fig.~2]{Macquart2020} 
is  $\sim$1000\,pc cm$^{-3}$,
and the lower limit on their  90$\%$-confidence interval is $\sim$700\,pc cm$^{-3}$,
we set the lower limit in our MCMC parameter search space to be 600\,pc cm$^{-3}$.
We discuss a number of implications from our simulation for the DM$_{\mathrm{IGM}}$
model in Sect.~\ref{Sec: discussion}.

\subsection{The CHIME/FRB detection system model}

\begin{figure*}[tbh]
  \centering
	\includegraphics[width=0.9\linewidth]{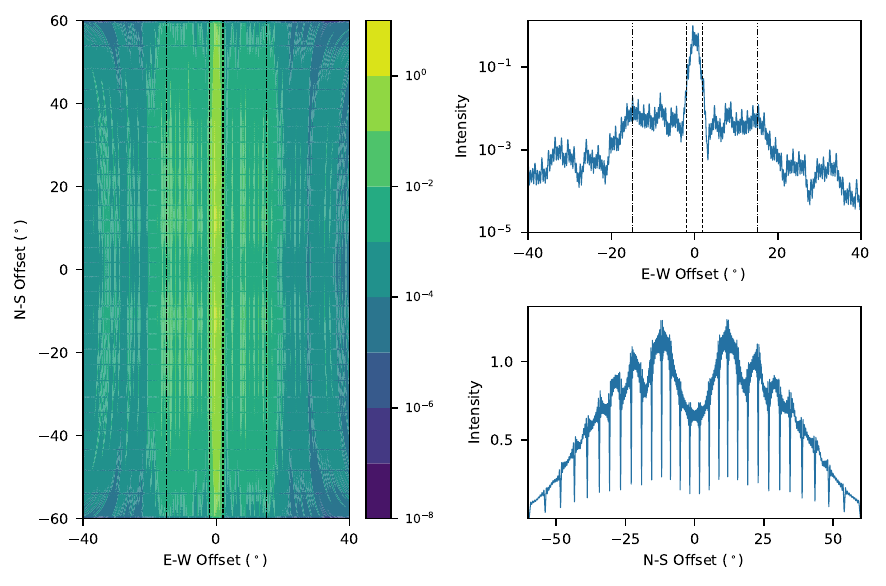}
    \caption{CHIME/FRB beam model reproduced with $\texttt{cfbm}$ at 600\,MHz. The left panel shows the beam
      intensity map in the range $[-40^\circ, 40^\circ]$ (East-West) and $[-60^\circ, 60^\circ]$ (North-South). The
      upper and bottom right panels show the East-West slice (y=0$^\circ$) and North-South slice along the meridian
      (x=0$^\circ$) respectively.
      The relative beam intensity is dimensionless, and normalized to the transit of Cyg A \citep{Amiri2022}.     
      To include the side-lobe in our simulation, the beam range $[-15^\circ, 15^\circ]$ (East-West) $\times\, [-60^\circ, 60^\circ]$ (North-South) is considered, where the East-West borders are indicated with dash-dotted lines. }
    \label{Fig: CHIME Beam}
\end{figure*}

Using \frbpoppy, we first generated simulated intrinsic populations according 
to certain parameter distribution models. We then set up
 ``surveys'' with survey parameters for specific telescopes. We applied each survey to each intrinsic
population, selecting the events that meet the  S/N threshold to form detected populations. We compared the simulated detected populations to the actually
detected population. From the best match, we inferred the actual intrinsic population \citep[as in][]{Gardenier2019}. 
A straw man CHIME survey was previously included in {\frbpoppy} \citep{Gardenier2021a}.
In Sect.~\ref{sec:sn_gain} we evaluate  the actual sensitivity of the telescope compared to these estimates.
Furthermore, in this work we used the implemented number of channels (16k), and an improved beam model.

\subsubsection{The beam model}
\label{sec:beam_model}
The earlier, initial approximation of the CHIME primary beam in  {\frbpoppy}
 consisted of a North-South cosine function convolved with an East-West Airy disk pattern.
The behavior of the synthesized beams formed in subsequent stages is more straightforward, and well known.
Since then, a CHIME primary-beam pattern measurement was empirically derived from the telescope response to the Sun \citep{Amiri2022}.
Using the accompanying \texttt{cfbm} package\footnote{\url{https://github.com/chime-frb-open-data/chime-frb-beam-model}},
we now included this improved CHIME/FRB beam intensity at 600\,MHz in  {\frbpoppy}.
The implemented beam model is
shown in Fig.~\ref{Fig: CHIME Beam},
left panel, with the East-West slice ($y$=0$^\circ$) and North-South slice along the meridian ($x$=0$^\circ$) shown in upper right and bottom right panels respectively.

Although only three one-off FRBs are detected outside the main beam (|East-West offset| > 2.0$^\circ$) in CHIME/FRB Catalog 1, 
this side-lobe detection fraction is important to distinguish models with different redshift distributions and luminosity models. 
The main-lobe and side-lobes arguably deliver both a deep and a shallow survey. 
Therefore, we did not restrict ourselves to the main beam but we consider the East-West range of [$-15^\circ$, $15^\circ$]. 
We end before the significant drop in intensity beyond that range. These borders are indicated with dash-dotted lines in
Fig.~\ref{Fig: CHIME Beam}. Although the far side-lobe of the CHIME/FRB beam is yet poorly understood
\citep{CHIMEFRB2021} and the three side-lobe events (FRB\,20190210D, FRB\,20190125B, and FRB\,20190202B) are excluded in many statistical studies, they are included in our simulations, as we expect them to 
put significant constraints on the number density models, luminosity distributions and high S/N events.

\subsubsection{Modeling CHIME surveying}
\label{Sec: modeling chime surveying}
In the surveying step, the S/N of an FRB is
derived from its peak flux density $\bar{S}_{\text{peak}}$ and pulse width.
This peak flux density \citep{Lorimer2013,Gardenier2019} is
\begin{equation}
    \label{Eqn: S_peak}
    \bar{S}_{\text {peak }}=\frac{L_{\mathrm{bol}}(1+z)^{\gamma-1}}{4 \uppi D(z)^{2}\left(v_{\text {high }}^{\prime \gamma+1}-v_{\text {low }}^{\prime \gamma+1}\right)}\left(\frac{v_{2}^{\gamma+1}-v_{1}^{\gamma+1}}{v_{2}-v_{1}}\right)\left(\frac{w_{\text {arr }}}{w_{\text {eff }}}\right),
\end{equation}
where $v_{\text{high}}^{\prime}=10$\,GHz, and $v_{\text{low}}^{\prime}=100$\,MHz, $v_2=800$\,MHz and $v_1=400$\,MHz are
adopted; $\gamma$ notates the spectral index; $D(z) = d_L(z)/(1+z)$ is the proper distance of the source; and the luminosity distance $d_L(z)$ is calculated with 
\begin{equation}
	d_L(z) = \frac{c (1 + z)}{H_0}\int_{0}^{z}\frac{\mathrm{d} z}{\sqrt{\Omega_m (1 + z)^3 + \Omega_\Lambda}},
\label{eqn:dL}
\end{equation}
where we used Planck15 results
$H_0$ = 67.74\,km\,s$^{-1}$ Mpc$^{-1}$, $\Omega_m$ = 0.3089 and $\Omega_\Lambda$ = 0.6911 for a
flat Lambda cold dark matter ($\Lambda$CDM) universe \citep{Planck2016}\footnote{The Hubble-Lema\^itre tension \citep{Riess2021,Hu2023}
between the local distance indicator and the cosmic microwave background (CMB) can result in up to 10\% discrepancy in
distances and the inferred redshifts from DM$_{\text{IGM}}$. The impact of this uncertainty is not investigated in
this work.}.
The S/N is derived using
\begin{equation}
    \mathrm{S} / \mathrm{N}=I\frac{\bar{S}_{\mathrm{peak}} G}{\beta T_{\mathrm{sys}}} \sqrt{n_{\mathrm{pol}}\left(v_{2}-v_{1}\right) w_{\mathrm{eff}}},
\label{eqn:sn}
\end{equation}
where $I$ is the beam intensity at the detection location, $G$ is the gain, $\beta$ is the degradation factor,
$T_{\text{sys}}$ is the total system temperature specific to CHIME/FRB, 
$n_{\text{pol}}$ is the number of polarizations, and $\nu_{1,2}$ are the boundary frequencies of the survey, respectively \citep{Lorimer2004}.
For the CHIME/FRB gain, degradation factor and total system temperature we 
updated the preliminary numbers used in \cite{Gardenier2021a}
and follow \citet{Merryfield2023}.
The measured system equivalent flux density (SEFD) reported there ranges from 30 to 80 Jy over the band. 
\citet{Merryfield2023} initially used a 
value of 45\,Jy in their injection system but find their idealized assumptions
do not represent the system adequately; requiring an increase of the injection
threshold from 9$\sigma$ to 20$\sigma$.
We here aim to include this real-life factor of 2 
over the theoretical performance for an SEFD of 45\,Jy
by using the average SEFD reported over the band
(55\,Jy, implemented as  gain $G$ = 1\,K\,Jy$^{-1}$ and $T_{\text{sys}}$ = 55\,K)
and a degradation factor $\beta$ of 1.6.
We discuss the influence of these updated numbers in Sect.~\ref{sec:sn_gain}.

Up to and including \frbpoppy 2.1 the S/N formula, Eq.~\ref{eqn:sn}, erroneously used $w_\text{arr}$
\cite[Eq.~17 in][]{Gardenier2019}, not $w_\text{eff}$, for the observed pulse width.
This meant that while  $\bar{S}_{\text{peak}}$ was correctly decreased by
the factor $w_{\text {arr }}/{w_{\text {eff }}}$
in Eq.~\ref{Eqn: S_peak},
the accompanying, partly counterbalancing
increase in S/N by $\sqrt{w_\text{eff}}$ in Eq.~\ref{eqn:sn} was not properly accounted for.
This made smeared-out pulses harder to detect than in real life.
That is corrected in \frbpoppy 2.2 and in our results below.
The impact of this change is, for example, that higher-DM FRBs are slightly easier to detect,
influencing $\alpha$. 
To determine the impact of the change,
we compared the best-fit Euclidean models against it,
and find that the values describing the underlying population
(Table~\ref{Tab: best-fitting parameters}) changed by about 0$-$0.4$\sigma$.

\subsubsection{Comparisons with the CHIME/FRB injection system}
Part of our approach is similar, in goals and method, to the CHIME/FRB injection system
mentioned above \citep{Merryfield2023}.
There, a mock population of synthetic FRBs was injected into the real-time search pipeline to
  determine the selection functions. Our main goal with \frbpoppy is to next go beyond this essential expression of the
  selection functions: we focus on determining the intrinsic population distributions, and we argue that the best way to include the relevant biases  imposed by the Universe and telescope is through forward modeling. 

In contrast to using 
the selection functions or fiducial distributions from \citet{CHIMEFRB2021}
or \citet{Merryfield2023}, 
the flexibility of forward modeling allows us to determine
which selection factors contribute most,
thus leading to better understanding of the interaction between the intrinsic population and the telescope strengths and weaknesses.
This is only possible if the survey is modeled in the same simulation as the population. 
Furthermore, certain parameters --  e.g., the spectral index, the activity dependence on frequency -- 
can only be done in a multi-survey simulation, which means one has to treat selection effects similarly for all.

\section{Methods}
\label{Sec: methods}

Our aim is to find the global best-fit model for describing the FRBs emitted in the Universe,
which are then input to the modeled telescope.
The sample of bursts can be well described by of order 10 characteristic numbers.
A fit over such a large number of parameters is, however, computationally challenging.
To realize this after all, we have improved {\frbpoppy} code and application in three ways.
We first added {MCMC} sampling,
we next employed data reuse where possible, and we finally deployed  {\frbpoppy} on a supercomputer.
These are described below.

\subsection{Markov chain Monte Carlo}

\begin{figure}
	\centering
	\includegraphics[width=1.0\linewidth]{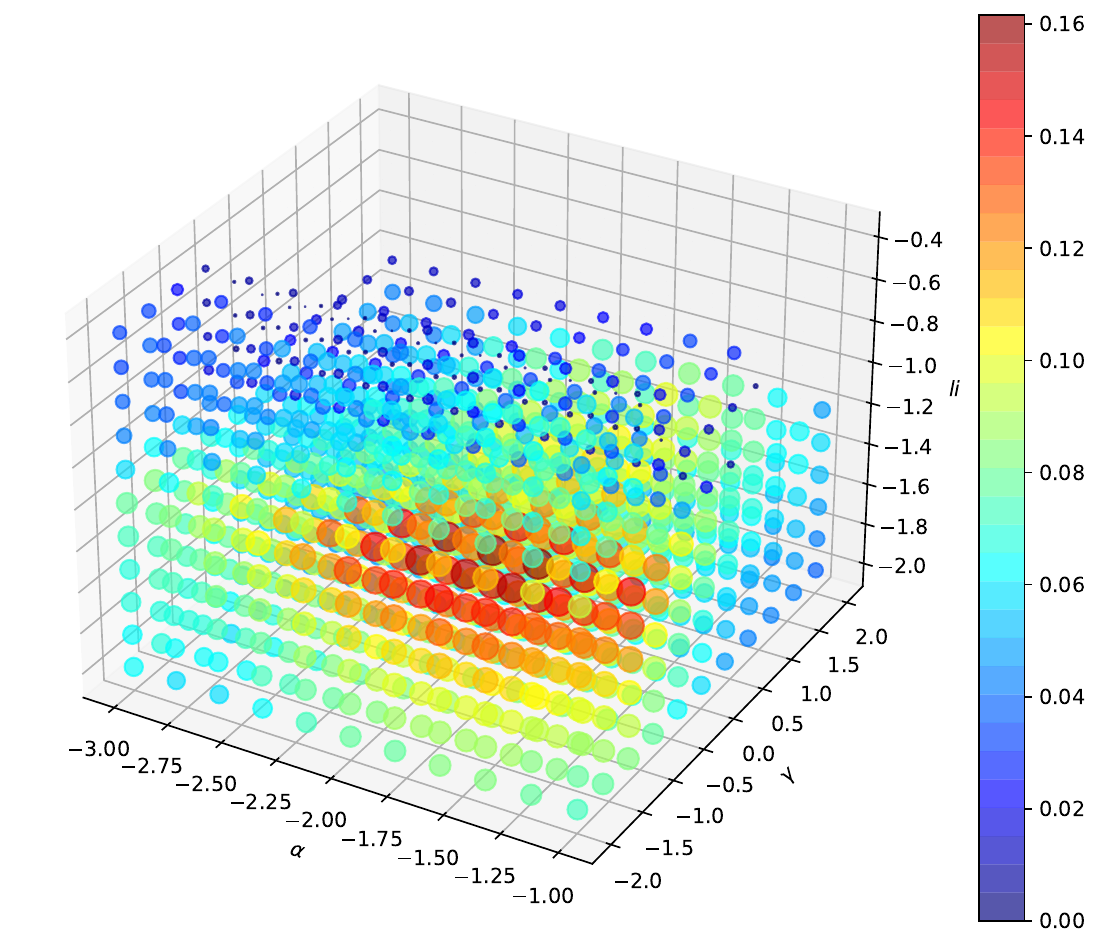}
	\caption{Three-dimensional GoF plot for \{$\alpha, \gamma, li$\} from an MC simulation. Different GoFs are denoted with colors as well as marker sizes.}
	\label{Fig: MC-alpha}
\end{figure}

\begin{figure*}[tbh]
\centering
\includegraphics[width=1.0\linewidth]{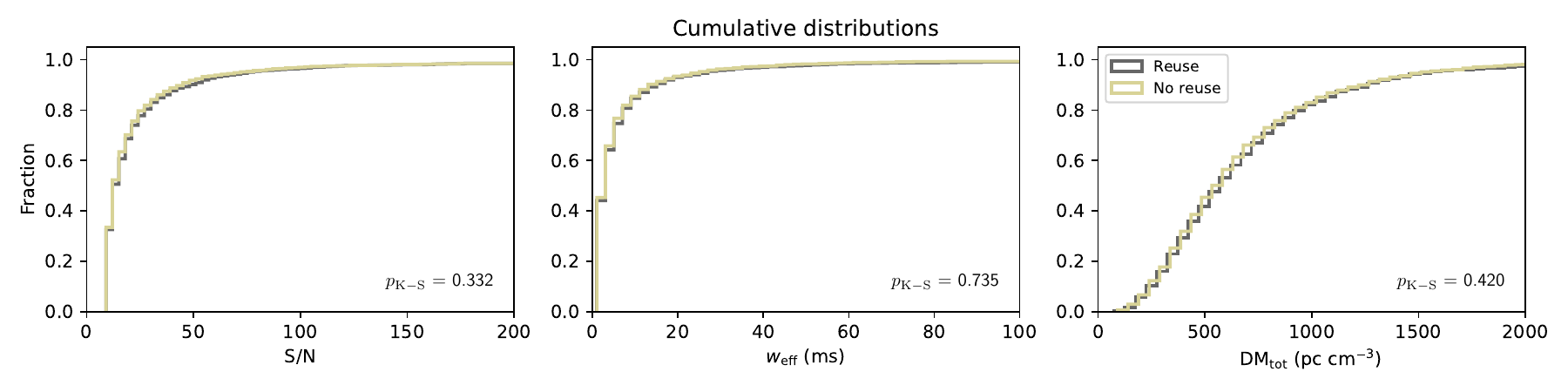}
\caption{Cumulative distribution plots of S/N, $w_\text{eff}$ and DM$_\text{total}$ distributions from two surveyed populations
  generated with and without reuse, for the Euclidean model. The $p$-value of K-S test is also shown for reference.}
\label{Fig: reuse}
\end{figure*}

\citet[][\citetalias{Gardenier2021b}]{Gardenier2021b} conducted
a multi-dimensional MC simulation over 9 parameters,
all explained above: $\alpha, \gamma\footnote{Called $si$ in  \citetalias{Gardenier2021b}}, li, L_\text{min},
L_\text{max}, w_\text{int, mean}, w_{\text{int, std}}, \text{DM}_{\text{IGM, slope}}, $
and
$ \text{DM}_{\text{host}}$. Due to computational limitations, in that work these
these 9 parameters were divided over 4 subsets, chosen to be maximally independent of each other,
but with a few shared parameters between sets,
such that the globally best model could arguably be approached:
1: \{$\alpha, \gamma, li$\}, 2: \{$li, L_\text{min}$, $L_\text{max}$\}, 3: \{$w_\text{int, mean}$,
$w_{\text{int, std}}$\} and 4: \{$\text{DM}_{\text{IGM, slope}}$, $\text{DM}_{\text{host}}$\}.
Each time, one subset was searched, keeping other parameters fixed,
using the best-fitting values of previous run as the
input of the subsequent run.
This was  repeated over multiple cycles
to approach the global optimum. Fig.~\ref{Fig: MC-alpha} serves as an example
of the three-dimensional goodness-of-fit (GoF) plot from such an {\frbpoppy} MC simulation;
in this case  for \{$\alpha, \gamma, li$\}.
A down side of this approach,
also discussed in \citetalias{Gardenier2021b},
is that during every run the uniform sampling means regions of low and high GoF are treated equally,
and searched with the same step size. Thus, the MC simulation spends
a considerable amount of computing time in regions that are not actually interesting.

Therefore, we implemented and conducted a full dimensional {MCMC}
simulation\footnote{Using the python module \texttt{emcee} \citep{Foreman-Mackey2013}, \url{https://github.com/dfm/emcee.}},
to supersede the \citetalias{Gardenier2021b} MC simulation with divided subsets.
As the MCMC sampler moves out of regions of poor GoF more quickly
than the brute-force method employed earlier,
this allows us to sample the full multi-dimensional
search space.
Additionally beneficial is that such sampling allows for error estimates on the outcome values,
something the previous, subset implementation lacked. 

For a given set of input parameters, we first generated a small population of $10^6$ FRBs,
then followed these through  our CHIME/FRB representation,
and saved the detected FRBs to the surveyed population.
Like in  \citealt{Gardenier2019}, we used the terms
``surveyed'', ``observed'' and ``detected'' synonymously.
We repeated the process until we had enough FRBs (e.g., 1000) in the surveyed population. 
To limit the
computational time required when sampling poor regions, we also stopped if after a certain number of iterations we did not have enough detections.

\citetalias{Gardenier2021b}
use the $p$-value $p_{\text{K-S}}$ from the two-sample Kolmogorov-Smirnov (K-S) test as GoF in their
MC simulation, to evaluate which population is more similar to the observed population. 
The growing size of the observed sample, however, means the high S/N or high DM events are approaching the actual
boundaries of the population. Since the models or parameters we are trying to constrain make different predictions for
these boundaries, the outlier events can be of great significance to evaluate population models.
Therefore, we have
extended \frbpoppy with the option to use the $k$-sample Anderson-Darling (A-D) test  \citep{Scholz1987}
as the measure to determine the  GoF.
The statistic $A^2$ of the A-D test uses weighting functions when it sums the cumulative distribution function (CDF) distances between samples, giving more
weight to the tail of distributions. In maximum likelihood estimation (MLE), the log-likelihood function then is the negative of the loss function (the statistic)
\begin{equation}
    \ln L \propto - A^2.
\label{eq:LA2}
\end{equation}

To represent the main features of FRB populations, we compared the S/N, $w_\text{eff}$ and DM$_{\text{total}}$
distributions from a simulation with the CHIME/FRB Catalog 1 one-off FRBs\footnote{For S/N, we compare with \texttt{bonsai\_snr} in CHIME/FRB Catalog 1.}.
To combine these three  distributions  to constrain the parameter set, we used as the log-likelihood function $\ln L$  the sum of the three statistic $A^2_i$
\begin{equation}
\ln L = - A^2_{\text{S/N}} - A^2_{w_\text{eff}} - A^2_{\text{DM}}.
\label{eq:LA}
\end{equation}
Here, we assume that the three distributions are uncorrelated, and that when $\ln L_\text{max}$ is reached,
we have identified the joint  minimum of  $A^2_{\text{S/N}}$, $A^2_{w_\text{eff}}$ and $A^2_{\text{DM}}$.
We used $\ln L$ (i.e.,~$-\sum A^2_i$) as the total GoF to evaluate how well a  set of parameters within the specific model  reflects  the simulated samples match real observations.
We discuss these statistics more in Sect.~\ref{sub:stat}.

\subsection{Speeding up of \frbpoppy}
In order to feasibly operate  \frbpoppy within an MCMC, a number of
code optimizations and changes were implemented, to speed up the generation of large populations. 
We now use \texttt{numpy} array look-up methods, superseding the SQL database approaches used in
 \frbpoppy 2.1.0 \citep{Gardenier2019}, to query the $\text{DM}_{\text{MW}}$ and the cosmological distance $d_L$. With \texttt{numpy.searchsorted}, we can efficiently find the indices where elements (coordinates or redshifts) should be inserted into a sorted one-dimensional array and then use indexing to obtain the required $\text{DM}_{\text{MW}}$ or $d_L$

A population in \frbpoppy contains property values for e.g., distance, luminosity, pulse width and DM.
When generating a new population, \frbpoppy now has the option to reuse some of these quantities.
This is especially beneficial for values  that are
expensive to generate but are uncorrelated with other values. In this way,
run time can be reduced by $\sim$50\%. Under this mode, we started from a relatively small population ($10^6$ FRBs) for which all parameters are drawn from the selected model.
In the simulated population, the redshift $z$, distance $d_\text{co}$, coordinates ($g_l$, $g_b$), and dispersion
measure (including $\text{DM}_\text{host}$, $\text{DM}_\text{IGM}$ and $\text{DM}_\text{MW}$) are coupled. They should
be reused as a whole. On the other hand, the FRB-intrinsic properties, i.e., the luminosity $L_\text{bol}$,
and pulse width $w_\text{int}$ are independent of this first set,
and can safely be regenerate from the parent distribution. These create new FRBs that
are not meaningfully correlated with those in the genesis population. All small populations thus generated are next
merged to form a large population (hereafter, the combined population). This means that each set ($z$, $d_\text{co}$, $g_l$,
$g_b$, $\text{DM}_\text{host}$, $\text{DM}_\text{IGM}$, and $\text{DM}_\text{MW}$) is used $N$ times in the large
population, combined with different $L_\text{bol}$ and $w_\text{int}$.
To validate this approach,
we  generated surveyed populations both with and without reusing these quantities,
and find they are practically identical, with similar K-S test $p$-values (see, e.g., 
Fig.~\ref{Fig: reuse}). Fewer than 1\% of FRBs have identical  coordinates and redshifts.
As a further check, we compared full MCMC runs of the Euclidean model with and without reusing quantities, and found no significant
differences in the  best-fitting values.
Hence, the new strategy is robust and does
not influence our detection of FRBs. 

We employ this reuse and the other optimizations in the results discussed in the remainder of the paper.
Using these improvements, 
\frbpoppy 2.2.0  generates a population about 20 times faster than \frbpoppy 2.1.0.

\begin{figure*}
	\sidecaption
	\includegraphics[width=0.7\linewidth]{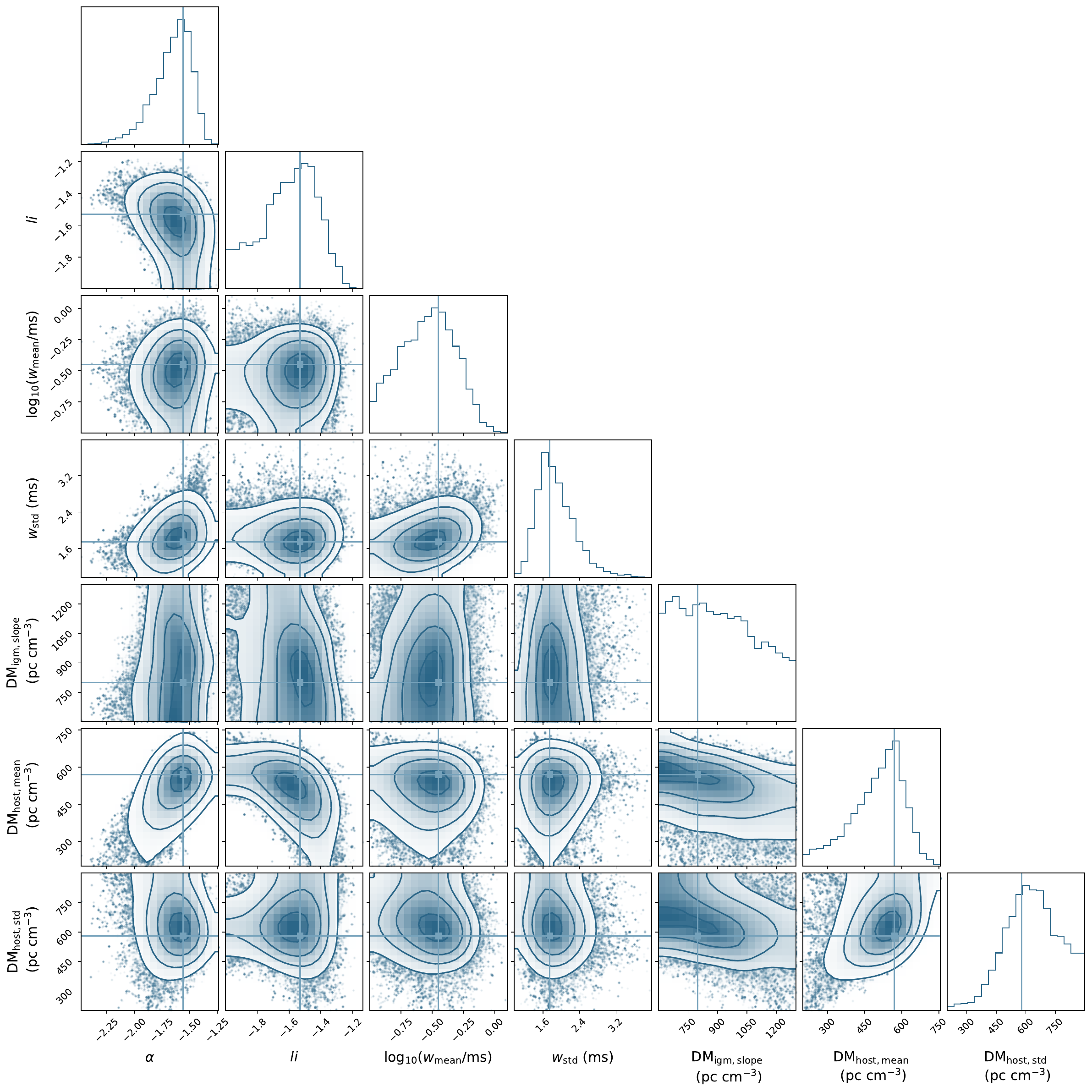}
     \caption{Confidence contours and marginalized likelihood distributions for the 7 parameters in our power-law number density model.}
	\label{Fig: MCMC-power-law}
\end{figure*}

\subsection{Computations}The computations were carried out on the Dutch national supercomputer
Snellius\footnote{\scriptsize\url{https://www.surf.nl/en/dutch-national-supercomputer-snellius}},
using the 128-core ``thin'' or ``rome'' nodes.
For our default MCMC simulation of 240 nwalkers $\times$ 500 steps, the computational cost
is 2$-$8 $\times$ 10$^4$ core hours, depending on the FRB population models and beam intensity map.

\section{Results}
\label{Sec: results}
Below we present the outcomes of our MCMC simulations. The interpretation is covered in Sect.~\ref{Sec: discussion}.

\subsection{Best-fitting parameters from different number density models}
\subsubsection{Power-law number density model}
\label{sec:results:pl}
The confidence contours and marginalized likelihood distributions for the power-law number density models are
shown in Fig.~\ref{Fig: MCMC-power-law} and the best-fitting values with 1$\sigma$ uncertainties are listed in
Table~\ref{Tab: best-fitting parameters}, for the population
parameter set
\{$\alpha, li, \log_{10} w_{\text{int, mean}}$, $w_\text{int, std}$, $\text{DM}_{\text{IGM, slope}}$, $\text{DM}_{\text{host, mean}}$, $\text{DM}_{\text{host, std}}$\}.

\begin{table*}[bth]
    \small
    \caption{Summary of best-fitting values for different models in this work and \citetalias{Gardenier2021b}.}
    \centering
    \def\arraystretch{1.5}
    \begin{tabular}{l|cccccccc}
    \hline
    \multirow{2}{*}{\diagbox{Models}{Parameters}} & $\alpha$ & $li$ & $\log_{10} (w_{\text{int, mean}}/\text{ms}) $ & $w_{\text{int, std}}$ & DM$_{\text{IGM, slope}}$ & DM$_{\text{host}}$ & DM$_{\text{host, mean}}$ & DM$_{\text{host, std}}$ \\
    &&&& ms & pc\,cm$^{-3}$ & pc\,cm$^{-3}$ & pc\,cm$^{-3}$ & pc\,cm$^{-3}$\\
    \hline
    \citetalias{Gardenier2021b} & $-2.2$ & $-0.8$ & $-2.44$ & 0.6 & 1000 & 50 & - & - \\
    This work: &&&&&&\\
    Power law density & $-1.58_{-0.27}^{+0.09}$ &
    $-1.51_{-0.28}^{+0.09}$ &
    $-0.45_{-0.30}^{+0.15}$ &
    $1.75_{-0.26}^{+0.55}$ &
    $780_{-94}^{+349}$ &
     - & 
    $560_{-190}^{+37}$ &
    $610_{-124}^{+176}$ \\
    SFR & - & $-1.58_{-0.18}^{+0.10}$ &
    $-0.50_{-0.19}^{+0.22}$ &
    $1.65_{-0.28}^{+0.40}$ &
    $840_{-140}^{+303}$ &
     - & 
    $490_{-129}^{+74}$ &
    $520_{-134}^{+209}$ \\
    Delayed SFR - 0.1\,Gyr & - & $-1.64_{-0.23}^{+0.11}$ &
    $-0.52_{-0.28}^{+0.17}$ &
    $1.70_{-0.27}^{+0.31}$ &
    $790_{-104}^{+353}$ &
     - & 
    $510_{-103}^{+78}$ &
    $540_{-111}^{+189}$ \\
    Delayed SFR - 0.5\,Gyr & - & $-1.62_{-0.25}^{+0.10}$ &
    $-0.53_{-0.27}^{+0.19}$ &
    $1.68_{-0.25}^{+0.35}$ &
    $800_{-100}^{+352}$ &
     - & 
    $510_{-115}^{+68}$ &
    $550_{-121}^{+178}$ \\
    Delayed SFR - 1\,Gyr & - & $-1.65_{-0.22}^{+0.13}$ &
    $-0.54_{-0.28}^{+0.20}$ &
    $1.68_{-0.31}^{+0.33}$ &
    $840_{-140}^{+303}$ &
     - & 
    $510_{-111}^{+76}$ &
    $560_{-131}^{+169}$ \\
    \hline
    \end{tabular}
    \tablefoot{The best-fitting values of the population  parameters, plus and minus the 1$\sigma$ uncertainty, from
      the MC simulation in \citetalias{Gardenier2021b} and the MCMC simulations of the power-law density,
      SFR and three delayed
      SFR models in this work are presented. For 
      formatting unity,
        insignificant digits are also included in certain values and errors.
    }
    \label{Tab: best-fitting parameters}
\end{table*}

\begin{figure*}
	\sidecaption
	\includegraphics[width=0.7\linewidth]{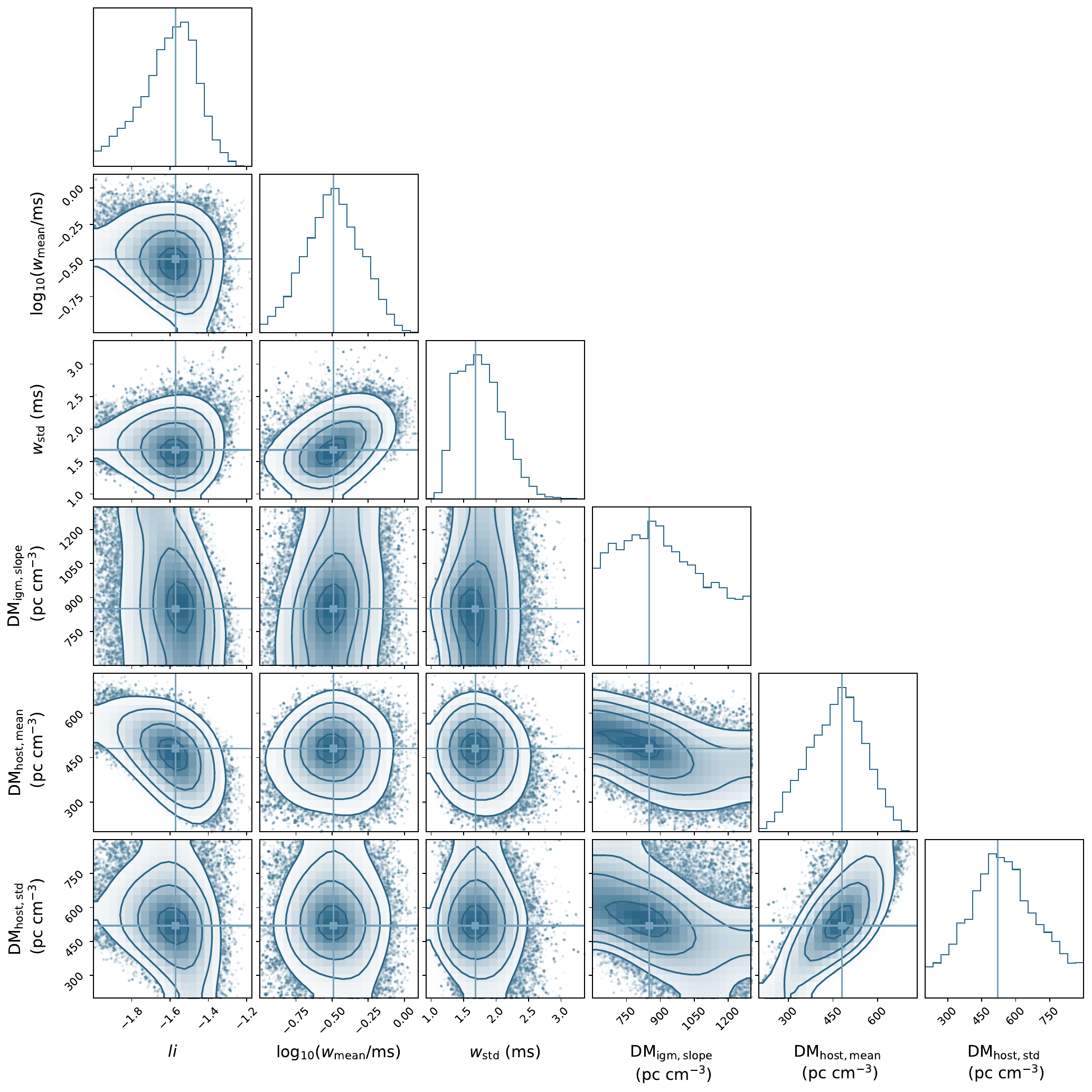}
 \caption{Confidence contours and marginalized likelihood distributions for the 6 parameters in our SFR
          model. }
	\label{Fig: MCMC-sfr}
\end{figure*}

\begin{figure*}
\centering
\includegraphics[width=0.8\linewidth]{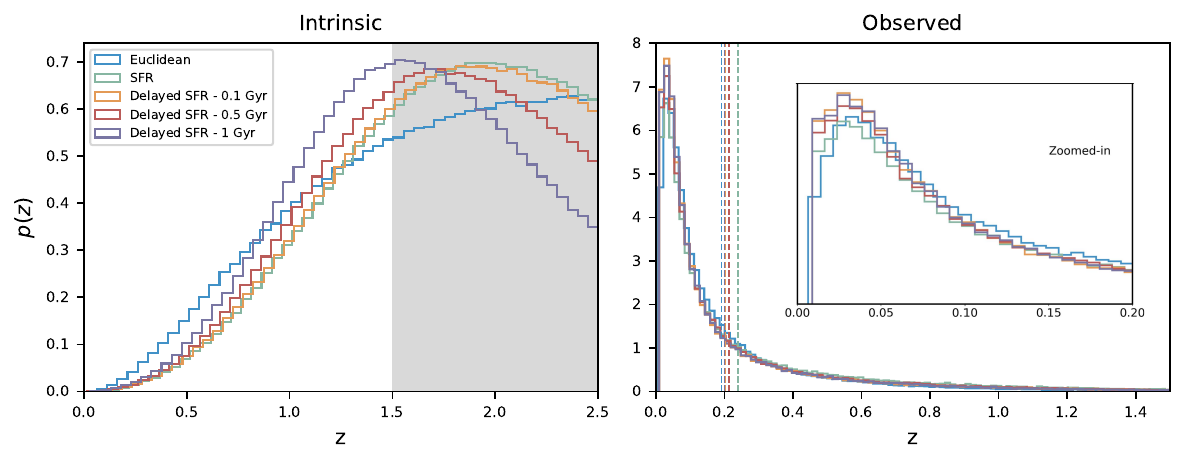}
\caption{Probability distributions of FRBs as functions of redshift for different models. 
\textit{Left panel}: The probability density functions (PDFs) of the intrinsic FRB population versus redshift, for the Euclidean, SFR and delayed SFR models with three delay times. The redshift range [0, 1.5] is used in our simulation. For information, we also show the region [1.5, 2.5], gray shaded, such that the shift of the SFR peak is visible between the models. 
\textit{Right panel}: The simulated observed FRB population PDFs against redshift, for the same models. The mean redshift <$z$> for the different models are marked by the dashed vertical lines. An unfilled zoomed-in view of the redshift range [0, 0.2] is also shown.}
\label{Fig: Number-redshift distribution}
\end{figure*}

While the number density power-law index $\alpha$ was $-$2.2
in the best overall model from \citetalias{Gardenier2021b},
we now find  $-1.58_{-0.27}^{+0.09}$ from the full MCMC.
That value is consistent with the non-evolving Euclidean distribution
\citep[cf.][]{Oppermann2016,James2019}.
We  thus hereafter refer to this best-fit power-law density model as the Euclidean model.
The luminosity index $li$ is constrained to
$-1.51_{-0.28}^{+0.09}$; and we discuss this finding, and compare it with other results -- minding the different
 definitions -- in Sect.~\ref{sec:res:lum}.
For pulse width, the best-fitting values of
$\log_{10} (w_\text{int} /\text{ms})$ is $-0.45_{-0.30}^{+0.15}$.
This produce mean intrinsic pulse widths two orders of magnitude higher than in
\citetalias{Gardenier2021b} -- this difference is explained in Sect.~\ref{Sec: modeling chime surveying}.

For the components that contribute to the  DM,
we find the $\text{DM}_{\text{IGM, slope}}$ is $780_{-94}^{+349}$\,pc\,cm$^{-3}$ while the
$\text{DM}_{\text{host, mean}}$, $\text{DM}_{\text{host, std}}$ are $560_{-190}^{+37}$\,pc\,cm$^{-3}$ and
$610_{-124}^{+176}$\,pc\,cm$^{-3}$ respectively.
We note here that the mean and standard deviation of $\text{DM}_{\text{host}}$ describe a log-normal distribution;
hence, they are quite far away from the more meaningful median value.
After conversion, the median of the observed  distribution for  $\text{DM}_{\text{host}}$ is
$\sim$330\,pc\,cm$^{-3}$ and its probability density function (PDF) peaks at $\sim$130\,pc\,cm$^{-3}$ (both in the source frame).
These values are larger than what \cite{Zhang2020} find from the
{\tt IllustrisTNG} simulation and \cite{Yang2017} from a sample of 21 FRBs. It is worth noting that the
$\text{DM}_{\text{src}}$ and $\text{DM}_{\text{MW, halo}}$ contributions are absorbed into the $\text{DM}_{\text{host}}$ term
in the simulations. $\text{DM}_{\text{MW, halo}}$ is not explicitly modeled in 
NE2001 \citep{Cordes2002,Cordes2003}. \cite{Prochaska2019} estimated
\mbox{$\text{DM}_{\text{MW, halo}}$ $\sim$50 $-$ 80\,pc\,cm$^{-3}$}. \cite{Yamasaki2020} reported a mean value of $\text{DM}_{\text{MW, halo}}$ $\sim$43\,pc\,cm$^{-3}$. For simplicity, in the following discussions and in Fig.~\ref{Fig: DM} in this work, we adopted a typical $\text{DM}_{\text{MW, halo}}$ of 40\,pc\,cm$^{-3}$ and subtracted it from the reported $\text{DM}_{\text{host}}$ results.

\subsubsection{SFR model}
\label{Sec: SFR}
We described the SFR model using parameters similar to the power-law density model (above)
but without requiring a number density power-law index $\alpha$, leading to the 6-parameter set
\{$li, \log_{10} w_{\text{int, mean}}$, $w_\text{int, std}$, $\text{DM}_{\text{IGM, slope}}$, $\text{DM}_{\text{host,
    mean}}$, $\text{DM}_{\text{host, std}}$\}. The confidence contours and marginalized likelihood distributions are
shown in Fig.~\ref{Fig: MCMC-sfr} while he best-fitting values with 1$\sigma$ uncertainties are also listed in
Table~\ref{Tab: best-fitting parameters}. The constraints on \{$li, \log_{10} w_{\text{int, mean}}$, $w_\text{int, std}$\}
are very close to those of the Euclidean model;
however, noticeable differences exist in \{$\text{DM}_{\text{IGM, slope}}$,
  $\text{DM}_{\text{host, mean}}$, $\text{DM}_{\text{host, std}}$\}. The median of observed source frame
  $\text{DM}_{\text{host}}$ is $\sim$279\,pc\,cm$^{-3}$ and its PDF peaks at $\sim$105\,pc\,cm$^{-3}$. This is larger
  than the median value 179 $\pm$ 63\,pc\,cm$^{-3}$ found in \cite{Mo2023} for their SFR model.

  In Fig.~\ref{Fig: MCMC-sfr}, the slightly diagonal confidence contours in the subpanels that 
  project $\text{DM}_{\text{IGM, slope}}$ against the  $\text{DM}_{\text{host}}$ parameters
  indicate
  that the {DM} contributions of the IGM and the host are somewhat degenerate with each other.
  This is because  the DM excess ($\mathrm{DM}_{\mathrm{total}} - \mathrm{DM}_{\mathrm{MW}}$)
  is supplied   by  the combination of  $\mathrm{DM}_{\mathrm{IGM,slope}}\,z$ and $\mathrm{DM}_{\text {host}}/(1 + z)$.
  Models that produce fewer local FRBs,  like the SFR  model,
   thus require a larger $\text{DM}_{\text{IGM, slope}}$ and a smaller $\text{DM}_{\text{host}}$
  to fit the observed distributions, compared to  the Euclidean model.

\begin{figure*}
\sidecaption
\includegraphics[width=0.70\linewidth]{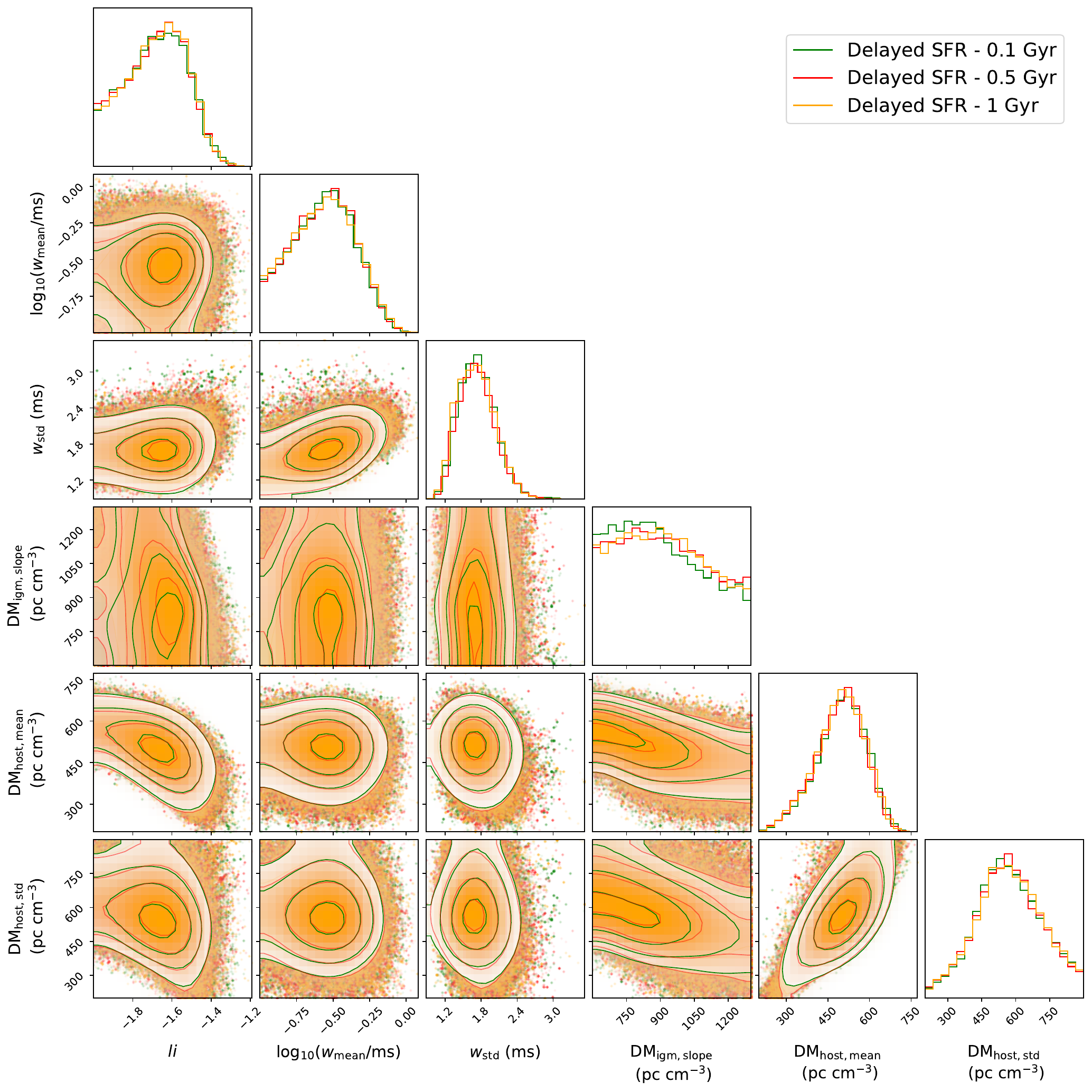}
\caption{Confidence contours and marginalized likelihood distributions for three delayed SFR model (0.1, 0.5, 1.0\,Gyr).}
\label{Fig: MCMC-delayed-SFRs}
\end{figure*}

\subsubsection{Delayed SFR model}
\label{Sec: res:delayedSFR}
To investigate the impact of formation channels that are delayed with respects to the SFR,
we simulated models with three different delay times, of 0.1\,Gyr, 0.5\,Gyr and 1\,Gyr.
In these models, the number of present-day, local FRBs is determined by the SFR when the delay process commenced.
As the recent SFR declines steeply \citep[see e.g.,][]{Gardenier2021a}, the number of local FRBs is expected to increase
with longer delay times. FRBs with a delay time of 1\,Gyr were formed near $z\simeq0.1$, where the SFR was
$\sim$30\% higher.
Thus, models with longer delay times in principle contain more low-$z$ FRBs; or, reciprocally, such models
require a lower source density to produce the number of FRBs that are observed.
As the simulated detection numbers in 
\frbpoppy are scaled to the actually detected number, we cannot easily identify an overall increase in the
number of local FRB sources. 
The GoF would only be affected if there was a significant change in slope within the sampled redshift range.
This effect is visible in the zoomed-in plot in the right panel of Fig.~\ref{Fig: Number-redshift distribution},
where the shapes of the three SFR models resemble one another very closely.
These are normalized to the same number of detections; the differences only emerge in the underlying physical rate. 
The 
posterior probability distributions 
are shown in Fig.~\ref{Fig: MCMC-delayed-SFRs}.
As all four SFR models have very similar best-fitting values, we cannot confidently distinguish between these.
We thus corroborate the conclusion from \citet{Shin2023} that there is insufficient evidence 
in CHIME/FRB Catalog 1
to strongly constrain how FRB evolution follows the SFR.
There is, however, so-called strong evidence \citep{Raftery1995}
in the Bayesian Information Criterion (BIC)
in favor of the  SFR models over the Euclidean model (Table~\ref{Tab: statistic}).
This conclusion
differs from \cite{Zhang2022}, who prefer a significant lag: a log-normal delay
model with a central value of 10\,Gyr and a standard deviation of 0.8\,dex.

\begin{table}[]
\caption{Summary of the A-D statistic and BIC for different models.}
    \centering
    %\linespread{1.5}
    \begin{tabular}{l|cccc|c}
    \hline
    \diagbox{Models}{Statistics} & $A^2_{\text{S/N}}$ & $A^2_{w_\text{eff}}$ & $A^2_{\text{DM}}$ & $A^2_{\text{total}}$ & BIC \\
    \hline
    Euclidean & 1.6 & 1.6 & $-$0.2 & 3.0 & 60.9 \\
    SFR & 1.1 & $-$0.3 & $-$0.1 & 0.7 & 48.4 \\
    Delayed SFR - 0.1\,Gyr & 1.3 & $-$0.6 & $-$0.8 & 0.2 & 46.8 \\
    Delayed SFR - 0.5\,Gyr & 1.1 & $-$0.6 & $-$0.6 & 0.2 & 46.8 \\
    Delayed SFR - 1\,Gyr  & 1.4 & $-$0.6 & $-$0.8 & $-$0.1 & 46.9 \\
    \hline
    \end{tabular}
    \tablefoot{
     These values are averaged from 50 realizations
      using best-fitting parameters for each model. The BIC shows no evidence in favor of the delayed SFR models over
      SFR model, and  strong evidence of over the Euclidean model. }
    \label{Tab: statistic}
\def\arraystretch{1}
\end{table}

\begin{figure}
    \centering\includegraphics[width=0.9\columnwidth]{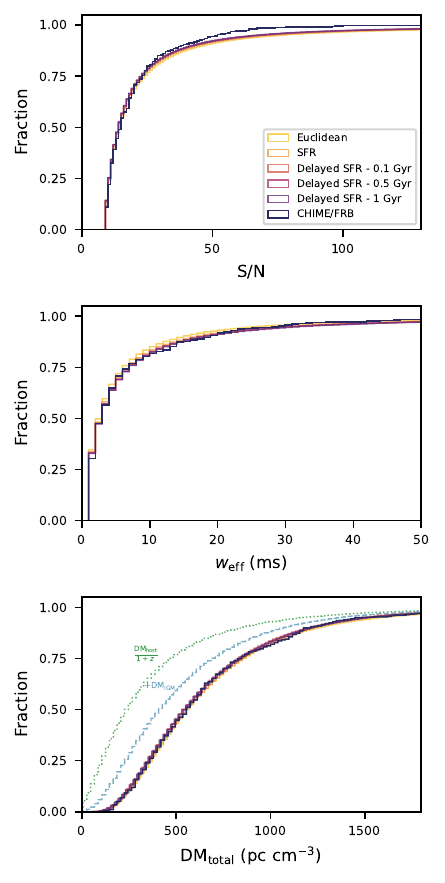}
    \caption{Comparison of cumulative distributions for simulated populations with best-fitting parameters and the
      CHIME/FRB Catalog 1 one-off FRBs. The S/N, $w_{\text{eff}}$ and DM$_{\mathrm{total}}$ are shown in upper, middle
      and lower panels respectively. The population coloring is the same in all subplots. In the DM plot we also show
      the cumulative contributions of $\text{DM}_{\text{host}}/(1+z)$, and of the additional $\text{DM}_{\text{IGM}}$,
      for the delayed SFR model with 0.1\,Gyr. These show how,
      for the lower-DM half of the population, these two components contribute roughly equally to the
      $\text{DM}_{\text{total}}$; while for the higher-DM half, the host contribution dominates.}
    \label{Fig: best fitting}
\end{figure}

\subsection{Reproducing CHIME/FRB distribution and model comparison}
With the best-fitting values from the MCMC, we reproduced the FRB populations and surveyed them with our model of CHIME/FRB. The CDF of S/N, $w$ and DM$_{\mathrm{total}}$ are shown in Fig.~\ref{Fig: best fitting}. The CHIME/FRB Catalog 1 distributions are also shown for comparison. The CDF curves of $w_\text{eff}$ show good agreements with CHIME/FRB while those of S/N has the largest discrepancy for all models. The overall good reproduction of all three distributions supports the method that we sum the A-D statistics to construct the log likelihood function.

To compare the models, we used the BIC (mentioned above):
\begin{equation}
    \mathrm{BIC} = k\ln n  - 2\ln L_{\text{max}}
\end{equation}
where $k$ is the number of free parameters in the model, $n$ is the number of samples.
In the power-law density models, $k=7$; in the SFR and delayed SFR models, $k=6$. A
summary of the A-D statistic and BIC for different models is provided in Table \ref{Tab: statistic}. \cite{Raftery1995}
list that BIC differences between 0$-$2, 2$-$6, and 6$-$10 correspond to ``weak'', ``positive'' and ``strong'' evidence
respectively. To avoid over-interpretation, we referred to BIC differences between 0$-$2
as only a hint of evidence.
Therefore, the three delayed SFR models, with almost the same BICs, show a hint of evidence over the SFR model;
and all four are strongly
favored over the power law density models
in general, and hence also over their most favored case, the Euclidean model.
The power law density model produces the least favorable fit, especially in the S/N distribution.

\section{Discussion}
\label{Sec: discussion}

Our results provide new insights into how many  FRBs are born in our local universe
and their evolution, into the brightness and emission of the bursts,
and into the baryonic material between the emitters and Earth. 
These warrant discussion and interpretation. We provide this below,
ordered from more general to more expert topics. 

\subsection{The number of FRBs in the local Universe}
\label{Sec:nFRBs}
Using \frbpoppy and the CHIME/FRB Catalog 1 we can establish of how many
FRBs are born and go off per day in our local
Universe, which we define to be up to $z=1$. 
We here consider FRBs purely as an observable phenomenon.
  Implications for any  underlying progenitor numbers are covered in the next subsection.
We present this fiducial number here to provide a best estimate based on our simulations.
The number depends on the model parameters (as discussed above)
and readers are invited to run their own simulations if desired. 
They can also directly download this snapshot as described in Sect.~\ref{sec:data_avail}.
We used our best-fit no-delay SFR model (with spectral index $-$1.5, luminosity index $-$1.58, and minimum luminosity
$10^{41}\,\text{erg}\,\text{s}^{-1}$).
We inputed the CHIME/FRB rate of $820\pm
60\,(\text{stat.})^{+220}_{-200}(\text{syst.})\,\text{sky}^{-1}\,\text{day}^{-1}$ for fluence > 5 Jy\,ms, scattering
time < 10\,ms at 600\,MHz, and DM > 100 pc\,cm$^{-3}$.

From these we determine the rate of  non-repeating FRBs that happen in the $z<1$ volume:  $10^{8.3\pm0.4}\,\text{day}^{-1}$. 
In other words, out to $z=1$, between $4\pm3\times10^{3}$ FRBs go off every second.

\subsection{Implications for the FRB progenitor volumetric rate}
\label{Sec:nProgn}

The FRB frequence determined in the previous subsection
  corresponds to an averaged volumetric rate density $\rho$ of  $2\times 10^8$ to $1\times 10^9$\,Gpc$^{-3}$\,yr$^{-1}$.
If one assumes FRB emission is concentrated with a beaming fraction $f_b$
(the fraction of the sky that is illuminated by the emitter beam),
then the FRB source rate is a factor 1/$f_b$ higher than this FRB rate. 
On the other hand, the input energy is emitted in a more focused way.
If we presume that the emission is uniform across the beam we can then convert the isotropic luminosity into a beamed
luminosity via $L'_\mathrm{bol} = L_\mathrm{bol, iso}/f_b$.
This increased luminosity means FRBs are  observable over a larger volume;
and this effect actually outperforms the sky illumination effects. In our simulations,
 lowering the beaming fraction thus reduces the required underlying source rate.

  Our rate is significantly higher than straightforward
  volumetric rates such as those quoted in \cite{Nicholl2017} as
$\rho \simeq 3.6 \times 10^4 f_b^{-1}$\,Gpc$^{-3}$\,yr$^{-1}$, and other early direct estimates.
These large apparent discrepancies arise because such work only takes the observed FRBs into account for their rates.
The aim in 
\texttt{frbpoppy} is to carry out  forward modeling, and correctly recover the intrinsic-to-observed selection, as shown
in Fig.~\ref{Fig: Number-redshift distribution}.
Only then can one aim to determine the underlying FRB formation rate from the telescope-observed rates.
The low-luminosity part of the underlying FRB population  dominates the rate, while these are only rarely detected.

The strength of \frbpoppy is its FRB population synthesis.
In future work, we will extend this to include progenitor populations;
while those who currently study certain  progenitor types,
could aim to produce the rates from Sect.~\ref{Sec:nFRBs}.
Still, a number of order-of-magnitude conclusions can already be drawn here.

We first investigate if double neutron star (DNS) mergers occur in sufficient numbers to produce our inferred FRBs.
If we assume the radio FRB  $f_b$ is similar to the high-energy beaming fraction,
then we can use typical $f_b$ = 0.04 found in short-duration gamma-ray bursts \citep[GRBs; ][]{Fong2015}.
The observed FRB rate would require a local volumetric DNS rate  
$\rho \simeq 6\times 10^7 $\,---\,$ 3\times 10^8$\,Gpc$^{-3}$\,yr$^{-1}$.
This far exceeds the 
local DNS coalescence rate of 1540$^{+3200}_{-1220}$\,Gpc$^{-3}$\,yr$^{-1}$
derived from GW170817 \citep{2017PhRvL.119p1101A}.
As a matter of fact, 
none of the  cataclysmic progenitor models suffice \citep[see, e.g.,][]{Ravi2019}.

As mentioned before, we aim to analyze the CHIME repeater FRBs in future studies;
but even if most apparent one-off FRB sources are very infrequent repeaters
such emission of multiple FRBs over the progenitor lifetime can greatly alleviate the rate problem.
\citetalias{Gardenier2021b} identify magnetars as the progenitors to FRBs. 
We first focus on younger magnetars, where the beaming fractions are high \citep[see e.g.,][]{2019A&A...623A..90S}.
Using the $f_b$ = 0.2 found for Galactic radio magnetars \citep{2012ApJ...744...97L}, % 3/13
we find an FRB emission rate of
$1  $\,---\,$ 7 \times 10^8$\,Gpc$^{-3}$\,yr$^{-1}$.
%%% 
If we take
an average supernova rate out to $z=1$ of $3 \times 10^{5}$\,Gpc$^{-3}$\,yr$^{-1}$ \citep{2012A&A...545A..96M}, a
magnetar fraction of $10^{-1}$ \citep{2008MNRAS.391.2009K}, % Keane & Kramer
and an activity lifetime of  $10^{3}$\,yr,
each source would need to produce $\mathcal{O}(10)$
 bursts per year -- almost all of which can be at the
low-luminosity end. We conclude this rate is possible.
If sources are older, longer-period magnetars \citep[cf.][]{2024arXiv240705366B},
the much larger active lifetime and the smaller beaming fraction work
 together toward even more attainable rates.

\begin{figure}[t]
\centering
\includegraphics[width=\columnwidth]{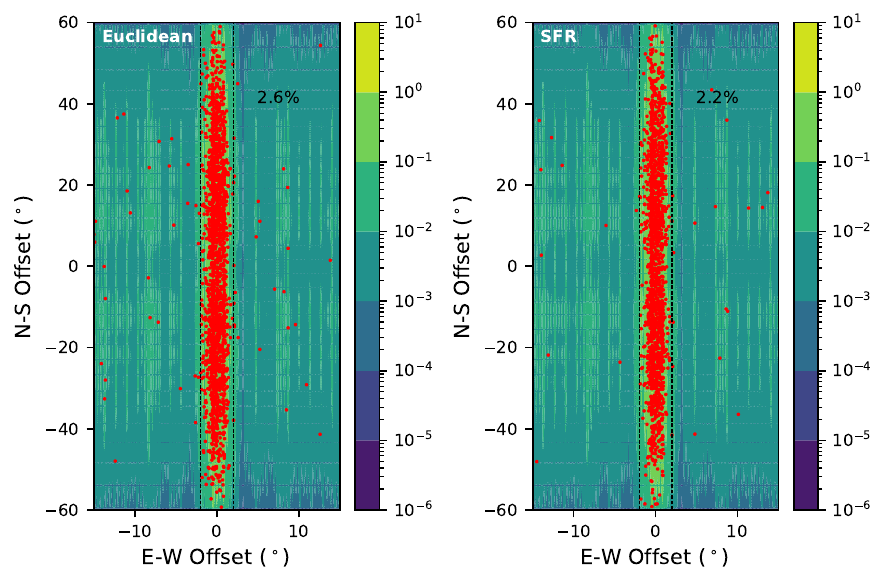}
\caption{Detection locations of simulated FRBs in the CHIME beam map (at 600\,MHz). The left and right panels are
  results of the Euclidean and {SFR} number density model respectively. The vertical dashed
  lines mark the chosen boundary ($\pm 2.0^\circ$) of the main-lobe. Side-lobe detections account for $\sim$$2.6\%$ and
  $\sim$$2.2\%$ in the  Euclidean and SFR models respectively, based on the average of 50 realizations. The beam intensity is normalized as in Fig.~\ref{Fig: CHIME Beam}.}
\label{Fig: location}
\end{figure}

\subsection{How FRB formation trails the SFR}
\label{sec:res:sfr}
In our models, the scenario where FRBs follow the delayed {SFR} is slightly favored over no delay (Table~\ref{Tab: best-fitting parameters}).
This agrees with the findings of \cite{James2022b}, but is in contrast with \cite{Zhang2022}.
As all SFR-based models are strongly favored over Euclidean models,
our results indicate that FRB emitters must be the direct descendants of a stellar population.
Source classes such as neutron stars, magnetars \citep{2020Natur.587...54C,2020Natur.587...59B},
and stellar mass black holes fit this description.
Coalescing neutron stars generally take a much longer time to merge \citep[e.g.,][]{2019ApJ...870...71P}.
Still we know of one FRB in an environment long devoid of star formation:
FRB~20200120E, in a globular cluster  of M81 \citep{2021ApJ...910L..18B,2022Natur.602..585K}.
That example means we should consider the possibility there is some fraction of FRBs that does exhibit a delay.
Proposed models for FRB~20200120E are induced collapse of a white dwarf through accretion ({accretion-induced collapse, AIC}),
or from the merger of two white dwarfs ({merger-induced collapse, MIC}).
In such models, the white dwarf inspiral is driven through gravitational-wave energy losses originally,
and finally through mass transfer.
\citet{2023arXiv230511933K} find that after 9\,Gyr, the approximate age of the M81 globular cluster,
white-dwarf mergers continue to occur, forming young neutron star that are the FRB emitters; although the relatively low
rate requires a long, 10$^5$\,yr active emitter lifetime.
Whatever the model,  FRB~20200120E exists and its formation trails the SFR peak by  billions of years.
It would be interesting to determine 
how many such FRBs can be mixed in with a general population of FRBs
that does closely follow the SFR.
To investigate this, we created a hybrid intrinsic population, of which 90\% of the FRBs follow SFR immediately,
while the other 10\% are  deferred by 1\,Gyr. Hence, these FRBs have a mean delay time of 0.1\,Gyr.
Although the hybrid population has more low-$z$ events,
it hardly deviates from the purely 0.1\,Gyr delay mode in our results. 
We conclude that the average delay time is the most important parameter.
As our models slightly favor short
delay times, the fraction of deferred FRBs cannot be large.
This can be quantified through the scenarios based
on the summary in Table~\ref{Tab: best-fitting parameters}.
For example,
if a fundamentally no-delay population contains a fraction $f_d$ FRBs formed with delay time $\tau_d$,
then the BIC  improves  by 1.6 if  $f_d \frac{\tau_d}{1 \mathrm{Gyr}} = 0.1$,
but then worsens again by 0.1 when $f_d \frac{\tau_d}{1 \mathrm{Gyr}} > 1$.

Other ways to elucidate the connections between FRBs and star formation  exist in principle.
From the locations of 8 FRBs within their host, \citet{2021ApJ...917...75M} find no convincing evidence
that FRBs strictly follow star-formation, nor that they require a delay;
and an FRB survey with Low Frequency Array (LOFAR) in star burst galaxy M82 finds no bursts \citep{2016A&A...593A..21M}.

\subsection{Side-lobe detection fraction}
\label{Sec: Side-lobe}
As discussed in Sect.~\ref{sec:beam_model},
CHIME effectively performs a deep (main-lobe) and shallow (side-lobe) survey
simultaneously.
The side-lobe plateau incorporated in our simulations ranges over $\pm15^\circ$ (East-West) and is
thus significantly larger than the main-lobe, as that roughly spans $\pm2^\circ$ 
(Fig.~\ref{Fig: CHIME Beam}).
On average, this side-lobe region under consideration is over a 100$\times$ less sensitive than the peak sensitivity.
Any FRBs detected there must be very bright;
and other such bright FRBs must then also occur in the main-lobe,
where they are seen as high-S/N bursts
\citep[see, also,][]{Lin2023}.
The 3 side-lobe FRBs reported in 
CHIME/FRB Catalog 1
have S/Ns of 21, 20, and 13. 
We would thus also expect $\sim$1\% of main-lobe FRBs to have S/Ns 100$\times$ higher, of 10$^3$ to 10$^4$.
And yet, the highest S/N detected in CHIME/FRB Catalog 1 is only 132.

We thus conclude that the CHIME/FRB pipeline missed a significant number of very bright FRBs.
We would expect these bursts to be relatively nearby and hence, display low DM.
\cite{Merryfield2023} mention
that the CHIME/FRB pipeline has a bias against such bright, low-DM FRBs through clipping during
 initial cleaning of RFI {radio frequency interference} at the L1 stage.
The detection efficiency suggested that $\sim$7\% of detectable injections above an S/N of 9 were mislabeled as RFI
\citep{Merryfield2023}.
Efforts to improve the CHIME/FRB pipeline in this regard could be worthwhile.
Bright, low-DM FRBs will be sources of great interest,
for exact localization within the host galaxy, multi-frequency
follow-up and for progenitor studies that help describe FRB formation.

\begin{figure*}
\centering
\includegraphics[width=1.0\linewidth]{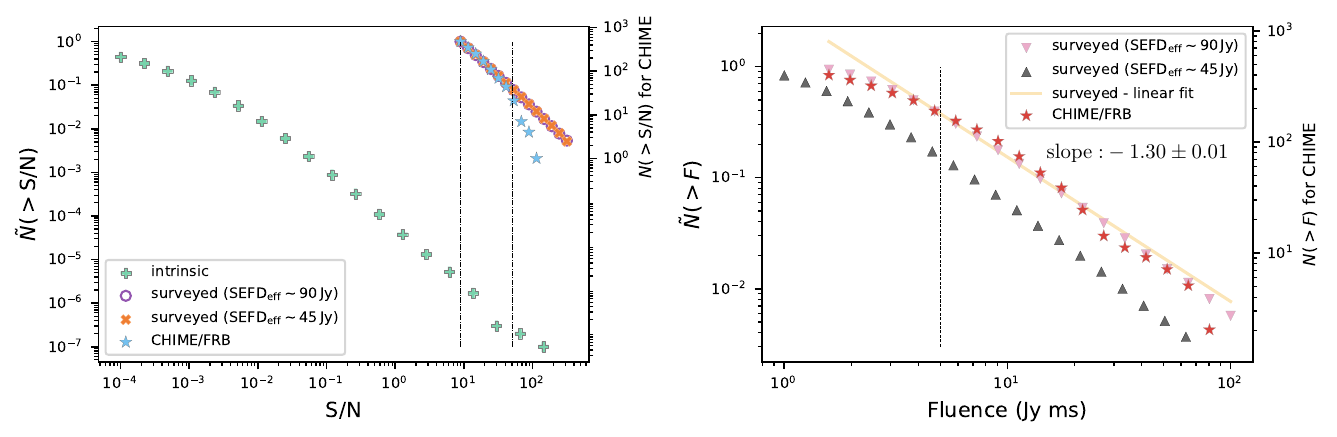}
\caption{Cumulative distribution of the S/Ns and fluences of FRBs in the SFR model. On the right ordinates,
  the number $N$ detectable/detected for CHIME.
  On the left ordinate, the fraction of this number over the total,  $\widetilde{N}=N/N_\mathrm{total}$.
  \textit{Left panel}: Plot of $N$(>S/N) versus S/N.
The two simulated  populations, surveyed with different $\text{SEFD}_\text{eff}$, are
shown with cross and circle markers, while the CHIME/FRB Catalog 1 is represented using star markers.
The S/N ``underlying'' distribution is generated using a threshold below the minimum range of this plot. The intrinsic and both surveyed populations have the same slope, indicating a
  scale-invariant feature despite selection effects and different telescope sensitivity. The vertical dash-dotted lines
  mark the S/N for the CHIME/FRB S/N threshold and for the start of the high-S/N deviation from our surveyed
  populations, respectively. \textit{Right panel}: $N$(>$F$)-$F$ plot. The surveyed population with different
  $\text{SEFD}_\text{eff}$ and CHIME/FRB Catalog 1 are shown in triangle down, triangle up and star markers
  respectively; the linear fit to surveyed population ($\text{SEFD}_\text{eff}\sim$ 90\,Jy has a slope $-1.30\pm
  0.01$. The vertical dashed line at 5\,Jy\,ms marks the lower bound used in our comparison of fluence distributions.}
\label{Fig: cumulative_SN_fluence}
\end{figure*}

Our simulations actually suggests that $\sim$2\% of the FRBs should have been 
brighter than the maximum observed S/N in Catalog 1 (discussed in more detail in the following Sections;
visible in Fig.~\ref{Fig: cumulative_SN_fluence}).
Local sources, originating from $z<0.1$, constitute $\sim$$60\%$ of those bright events.
Therefore, 
a rough estimate suggests that
CHIME/FRB may so-far have missed
of order 5 bursts
($\sim$1\% of its Catalog 1 size of $\sim$500;
\citealt{CHIMEFRB2021})
that are local,
bright FRBs -- these remain to  be discovered.

Our simulations indicate that equally bright FRB sources
must next also be detectable in the side-lobes more often than the reported  $< 1\%$ (3 out of 474).
We expect a side-lobe detection fraction of $\sim$$1.8\%$  in the delayed SFR models,
$\sim$$2.2\%$ in the straight SFR model, and $\sim$$2.6\%$ in the  Euclidean model.
The location of these detections are shown in Fig.~\ref{Fig: location}.
All the models considered here thus predict more side-lobe FRBs than Catalog 1 contains.
This prediction percentage is consist over a large number of simulations, but
the currently observed percentage could of course be affected by the small number statistics.

An alternative explanation for both absences is that there is some intrinsic fluence cut off, which prevents side-lobe
detections and high-S/N main-lobe detection alike.
We note here that this required fluence cut-off differs from a cut-off or change in the luminosity
  function. The latter is of course possible, and occurs at the emission point,
  where the luminosity is defined.
The fluence, however, is a value that is only defined at the
observer location. As it is determined by a combination of the distance to the source and its luminosity, which are
different for every FRB, the existence of such a cutoff seems very unlikely.
The more likely explanation appears to be
that there is a bias against high-S/N detections, and that some side-lobe detections go unrecognized as such.

\subsection{The CHIME sensitivity and its influence on the S/N and fluence distribution} 
\label{sec:sn_gain}
As introduced in Sect.~\ref{Sec: modeling chime surveying},
the initially predicted 
CHIME gain $G$ = 1.4\,K\,Jy$^{-1}$,
system temperature $T_\text{sys}$ = 50\,K and
degradation factor $\beta$ = 1.2
suggested in \citet{CHIMEFRB2018} and used in
\cite{Gardenier2019} are found
by the subsequent system study \citep{Merryfield2023}
to  overestimate the CHIME/FRB  sensitivity.
Therefore, in this work, we adopted $G$ = 1.0\,K\,Jy$^{-1}$, $T_\text{sys}$ = 55\,K, and $\beta$ = 1.6
to agree with that study.
These three parameters are most informatively combined when expressed as an effective
system equivalent flux density
SEFD$_\textrm{eff} = T_\text{sys}\beta/G$ (see Eq.~\ref{eqn:sn}). 
Thus, while \frbpoppy previously assumed SEFD$_\textrm{eff}$ = 45\,Jy, we now use SEFD$_\textrm{eff}$ = 90\,Jy.
This more realistic value underlies all above-mentioned results.

We here discuss the influence of this change on some of our metrics,
especially on the agreement of the observed and simulated S/N distributions (see, e.g., Eq.~\ref{eq:LA}).
In earlier sections, we have displayed these distributions as CDFs (e.g., Fig.~\ref{Fig: best fitting}).
In this current discussion, we display similar data but now as 
distributions of cumulative number $N(>$$x)$ versus $x$.
Observed sets of bursts, for both one-offs and repeaters,
are  commonly visualized this way for either the number $N(>$$x)$ or rate $R(>$$x)$,
against S/N, fluence or
luminosity as variable $x$ \citep[e.g.,][]{2020Natur.582..351C,Pastor2024}.

We find that the change in $\text{SEFD}_\text{eff}$
has no influence on the slope and GoF of the S/N distribution.
This indicates that despite selection effects and different telescope sensitivities,
the S/N distribution remains scale-invariant.
In Fig.~\ref{Fig: cumulative_SN_fluence} (left)
this is visible from the identical sets of points for the two surveyed sets.
In the same figure, the scale invariance is even more visible in the underlying 
S/N distribution (where we used  $\text{SEFD}_\text{eff}\sim$ 90\,Jy but no lower limit).  
The intrinsic population keeps a constant slope across almost the entire S/N range, thus validating the scale invariance
of the S/N distribution.
For comparison, the CHIME/FRB Catalog 1 sample is shown with star markers.
It diverges from our trend from S/N>50, suggesting CHIME missed high S/N events. 

While the S/N distribution cannot distinguish between  the old $\text{SEFD}_\text{eff}$ and the new, 
the fluence distribution can.
As shown in Fig.~\ref{Fig: cumulative_SN_fluence} (right)
the surveyed population with the updated  $G$ = 1.0\,K\,Jy$^{-1}$, $\beta$ = 1.6 and $T_\text{sys}$ = 55\,K
matches perfectly with the CHIME one-offs,
while the survey using the preliminary, more sensitive value has significant discrepancies.
For example,
38\% of FRBs in the CHIME/FRB Catalog 1
have fluences > 5 Jy\,ms (the dashed line in the right panel of Fig.~\ref{Fig: cumulative_SN_fluence}). The new  $\text{SEFD}_\text{eff}$  recreates this very well
(36\%) where the old one could not (17\%).
We are thus confident the degradation factor is made up of actually occurring (and normal)
 imperfections or inefficiencies in the survey, related to e.g., corrections for the beam intensities, or pipelines.

\subsection{Luminosity}
\label{sec:res:lum}
In the current study, the lower and upper boundary of the
luminosity are no longer free parameters as these are hard to constrain in the MCMC simulation. However, the
$L_\text{max}$ can influence the maximum S/N in the surveyed population. Continuing the discussion from the previous
subsections, the S/N threshold of CHIME/FRB Catalog 1 is 9
and the current maximum S/N is 132, which means the S/N range is smaller than two orders of magnitude. However, our
luminosity range spans five orders of magnitude (10$^{41}$ to 10$^{46}$\,erg s$^{-1}$). Under the power-law distribution
of luminosity, we would have expected 2\%, i.e., of order 10,
of the discovered FRBs to exceed this current maximum detected S/N of
132 (see Fig.~\ref{Fig: best fitting}, top panel). Of the detected sample, 0.1\% of FRBs should have S/Ns > 1000;  
Given that the  CHIME/FRB Catalog 1 contains  $\sim$500 FRBs, the non-detection of such a burst is not constraining.  

An exponential cutoff power distribution or smaller $L_\text{max}$ may result in a
smaller maximum S/N in the simulated sample,
although the proximity of the sources is unaffected, which is arguable more influential for this maximum S/N.
We thus still expect that many large-S/N FRBs are observed when the sample size grows. The maximum S/N can provide an important clue on the luminosity distribution
(e.g., a power-law or cutoff power-law, the $L_{\text{max}}$ or $L_{\text{cutoff}}$).  

Below, we compared our results for the  best-fit FRB luminosity models to other observational and theoretical results.
While we discuss these,
it is important to reiterate that two different definitions of the power-law index are commonly used,
and care should be taken when comparing the results. 

The luminosity power-law index $li$ in \frbpoppy is applied as $\mathrm{d} N(L)/\mathrm{d}\,\mathrm{log} L\propto
L^{{li}}$ or $\mathrm{d} N(L)/\mathrm{d} L\propto L^{{li}-1}$.
Under this definition, integration reproduces  $li$  as  the index in the cumulative distribution
$N(>L)\propto L^{{li}}$ of the generated population.
The other common definition is to report the exponent $\alpha$ from
above-mentioned  $\mathrm{d} N(L)/\mathrm{d} L\propto L^{\alpha}$,
as reported in e.g., \citet{2020Natur.582..351C}.
Thus, $\alpha$ = $li$$-$1. Both are generally negative numbers. So an $li$=$-1$ is identical to $\alpha$=$-2$.
The index  that determines the faint-end slope in the Schechter luminosity function
is defined as this same $\alpha$ \citep{Schechter1976}.
Our $li$ is this $\alpha$ plus one.

\begin{figure}
\centering
\includegraphics[width=\linewidth]{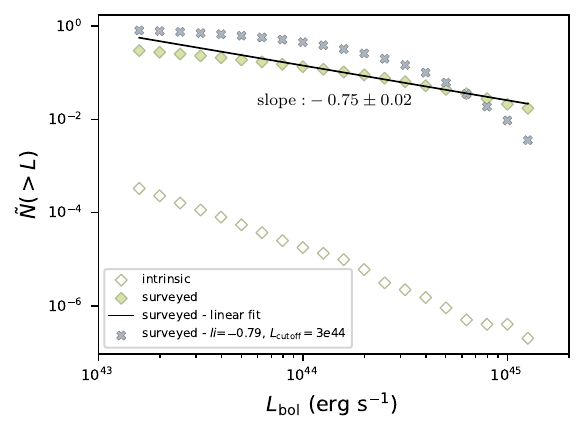}
\caption{Fractional cumulative number distribution of FRB luminosities,
  $\widetilde{N}(>$$L)$ against $L$,
  for the best-fit no-delay SFR model. The
  intrinsic population (hollow diamonds), follows the intrinsic luminosity index -1.58. 
  The surveyed, detected sample (filled diamonds), however, is modified by the survey selection function, 
  and follows a power-law with index $-0.75 \pm 0.02$ (fit result shown).
  Grey crosses indicate the surveyed population that would follow from the \citet{Luo2020} Schechter luminosity function.}
\label{Fig: cumulative_luminosity}
\end{figure}

Now, our $li$ could be measured directly as the slope in the log-log plot of $N(>$$L)$ versus $L$
of the intrinsic generated population.
Selection effects, of course, alter this slope for the observed cosmological population of one-off bursts.
This is displayed in Fig.~\ref{Fig: cumulative_luminosity}:
from an intrinsic population following $li$=$-$1.58,
few bursts with $L=10^{43}\,$erg s$^{-1}$ are detected (the curve flattens toward lower $L_\text{bol}$).
Brighter sources dominate the detected sample,
meaning the population CHIME sees follows the much shallower power-law with index $-0.75 \pm 0.02$.
But for observations of repeater bursts, where fewer such effects occur, this $li$ can be directly compared to the
fits (e.g.,  Fig.~4 in  \citealt{oostrum2020repeating} or
  Fig.~3 in \citealt{2024NatAs...8..337K}).

The preferred value for the luminosity index is $-1.58_{-0.18}^{+0.10}$ (Table~\ref{Tab: best-fitting parameters}).
That is interesting, because this value based on one-off FRBs agrees with
the value found by \citet[][there called $\mathit{\gamma}$]{oostrum2020repeating} for the repeating FRB~121102, of $-$1.7(6).
It is also consistent with the most complete,
high-fluence section of the broken power-law fit to  luminosity index for repeating FRB~20180916B 
reported as $-1.4\pm0.1$ in \citet[][there called $\mathit{\Gamma}$]{2021Natur.596..505P}.
This similarity in the pulse energy distribution between one-off and repeating FRBs,
something that would be unlikely to occur between completely difference sources classes,
corroborates the finding of \citet{Gardenier2021a}, recently confirmed by \citet{2023PASA...40...57J},
that the two apparent types both
originate from a single  population of FRB sources that is actually mostly uniform.

A number of other studies of one-off populations, in contrast, find different values for the luminosity index, but there
are important differences and caveats. We describe those below.

The study in \citet{Luo2020},
while possibly limited by the small sample size,
is similar in set up to {\frbpoppy}, albeit without simulating cosmological source evolution.
 \citet{Luo2020} assumed a Schechter luminosity function
and a log-normal intrinsic pulse width distribution, next applying a
flux threshold $S_\text{min}$ when calculating the joint likelihood function.
Their best-fit model based on 46 FRBs from 7
surveys is formed by  a Schechter luminosity index  $\alpha$ of
$-1.79_{-0.4}^{+0.3}$ with cutoff luminosity $\sim$$3\times$\,$10^{44}$\,erg\,s$^{-1}$.
This index, which corresponds to our $li$ = $-0.79_{-0.4}^{+0.3}$,
governs the faint end of the  $N(>$$L)$ distribution, up to the turn-over luminosity.
As FRBs with luminosities below $L=10^{43}\,$erg s$^{-1}$ do not significantly contribute to the observed population
(see Fig.~\ref{Fig: cumulative_luminosity}), the value of the index is not actually constraining.
It is the turn-over itself that determines the distribution. 
To illustrate this behavior
Fig.~\ref{Fig: cumulative_luminosity} also displays
the resulting surveyed population, 
from a one-off experiment with  a Schechter function in \frbpoppy. 
Although the distribution is less linear, the \citet{Luo2020}
result  is similar to our best model in overall average slope.

\cite{Shin2023} next report a Schechter index  $\alpha$ (called the differential power-law index there)
of $-1.3^{+0.7}_{-0.4}$, which in our notation corresponds to $li$ = $-0.3^{+0.7}_{-0.4}$ --
but for the energy, not the luminosity. 
That result, too, is based on  CHIME/FRB Catalog 1 data.
The modeling, however, is more limited.
Where \frbpoppy finds the best model by fitting over the  S/N, $w_\text{eff}$ and DM$_{\text{total}}$ distributions,
\cite{Shin2023} optimizes for the DM$_{\text{total}}$$-$Fluence distributions.
Fluence may be closely related to the luminosity index we discuss here,
but it can only be turned into S/N, the observed
quantity, through the pulse width.
In \cite{Shin2023} this width is drawn from the inferred observed width distribution,
not from an independent source distribution. 
To ensure understanding of the interplay between pulse width, luminosity and energy,
forward modeling of the intrinsic pulse width and any broadening effects from the intervening plasma have been part
of \frbpoppy since inception \citep{Gardenier2019}.
One of the main conclusions in \citet{Merryfield2023}, too, is that the CHIME S/N is affected much more 
strongly by the effective burst width than by the {DM}.
As the pulse widths are not explicitly  modeled in \cite{Shin2023},
the use of a single energy distribution, not
separate luminosity and width distributions, is required. 
A number of  inherent selection effects due to width may be absorbed into the energy function. 
We hence do not think a meaningful discussion of the astrophysics implied by our
different luminosity and energy results is possible. 

\cite{James2022a} find that the luminosity index dominates the extent to which FRBs are seen from the local universe, 
as suggested earlier by \cite{Macquart2018} --
with steeper values indicating more nearby events in the observations.
Therefore, the differences in the required $li$  reflect the different FRB
$z$ distribution:
\cite{James2022a} find $li$=$-1.09_{-0.10}^{+0.14}$ and in this case,
the fraction of FRBs  in the observed sample with $z<0.1$ is  $\sim$10\%; 
while in our best model, the more negative $li$=$-1.58$ requires a larger fraction of local FRBs in the surveyed sample,
of $\sim$50\% below $z=0.1$.

\subsection{Pulse width}

\begin{figure*}
    \centering\includegraphics[width=1.0\linewidth]{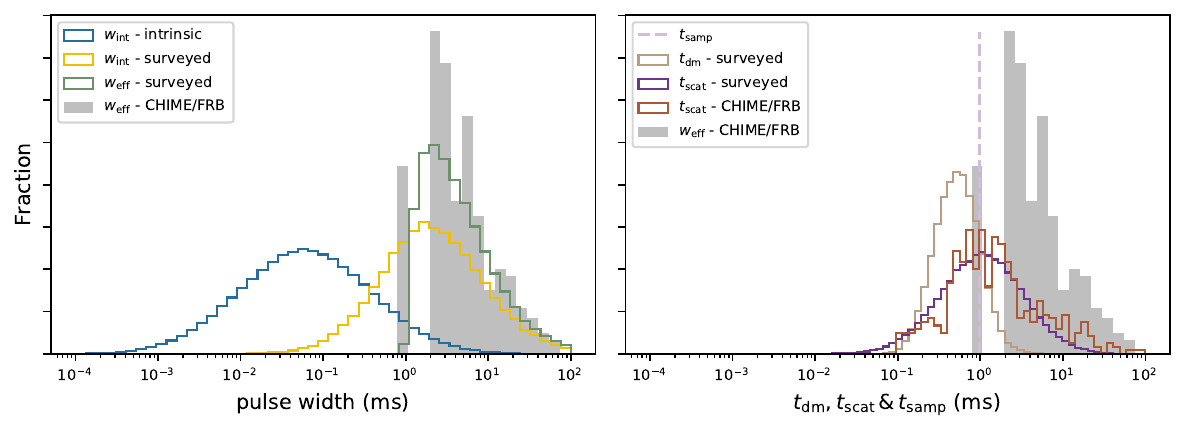}
    \caption{Probability distribution of pulse width and broadening effects. \textit{Left panel}: The normalized distribution of intrinsic pulse width $w_{\text{int}}$,
      for the intrinsic, generated sample and for the surveyed, detected sample,
      the effective pulse width $w_{\mathrm{eff}}$ of this surveyed sample in the simulation,
      and of the CHIME/FRB Catalog 1 widths. \textit{Right panel}: The
      distribution of the simulated normalized $t_\text{dm}$ and $t_\text{scat}$
      subcomponents of the width of the surveyed, detected FRBs; compared to the values from the
      CHIME/FRB Catalog 1. The sampling time $t_\text{samp}$ of 0.98304\,ms is marked with a dashed line. The population
      is generated using best-fitting values for the 0.1-Gyr delayed SFR model.}
    \label{Fig: pulse width}
\end{figure*}

The log-normal distribution of intrinsic pulse width is inspired by those observed in pulsars and repeater pulses \citep{Gardenier2019}. It can naturally produce bursts from microseconds to a few seconds FRB duration, and cover at least six orders of magnitude, which is hard to achieve with a normal distribution. 
The best-fitting values for the pulse width distributions 
are constrained in our MCMC simulation and are consistent among different models:
$w_{\text{int, mean}} \simeq 0.3$\,ms and $w_{\text{int, std}} \simeq 2$\,ms
  (Table~\ref{Tab: best-fitting parameters}).
The result is a relatively broad best-fit input width distribution (Fig.~\ref{Fig: pulse width}, left panel).
Bursts with larger intrinsic widths are also the higher fluence bursts.
That is why the high-width tail is preferentially detected.
Still, the pulse width contours in e.g., Fig.~\ref{Fig: MCMC-sfr} are relatively well localized,
indicating  high confidence that this distribution shape is preferred.
The slight correlation visible in the $w_{\text{int, mean}}$\,--\,$w_{\text{int, std}}$
projection in that figure, showing that larger widths are required at higher means,
indicates that generation of the sub-millisecond side is important.

The scattering time follows a log-normal distribution, as is visible in Fig.~\ref{Fig: pulse width} (right panel).
We compare our simulation outcomes against the boxcar width (bc width)  provided in the CHIME/FRB Catalog 1,
which takes an integer multiple of the 0.98304\,ms time resolution.
For FRBs of small intrinsic pulse width
$w_\text{int}$, the effective pulse width $w_\text{eff}$ is dominated by $t_\text{scat}$, $t_\text{dm}$ or
$t_\text{samp}$.
Fig.~\ref{Fig: pulse width} shows that although there are many events with $w_\text{int}$ below 1\,ms,
all of the $w_\text{eff}$ are above 1\,ms due to the scattering and instrumental broadening around 600\,MHz.
We note that for many FRBs, the CHIME/FRB
Catalog 1 only provides their $t_\text{scat}$ upper limits, which may cause a bias in modeling the lower end of
$w_\text{int}$.

We see here how two opposing selection effects are at play.
In contrast to other models, that are energy based \citep[and thus work in units of ergs; see][]{James2022b}, 
our model is luminosity based (e.g., working in units of erg\,s$^{-1}$).
This prevents simulations in which unphysically large amounts of energy are output per time unit.
Some FRBs are as short as 10$^{-6}$\,s
\citep[see, e.g., ][and note in Fig.~\ref{Fig: pulse width} that these are indeed created in \frbpoppy]{2023NatAs...7.1486S}, others as bright as
$10^{42}$\,erg \citep{2022arXiv221004680R};
allowing a model to combine these would result in luminosities over 
 $10^{48}$\,erg\,s$^{-1}$.
Such a luminosity exceeds
the maximum theoretical FRB luminosity under the assumptions of \citet{2019MNRAS.483L..93L},
and the maximum spectral luminosities allowed in all but the most extreme parameter combinations
in the models of \citet{2021MNRAS.508L..32C}. 
As described in Sect.~\ref{sec:lum}, the top end of our luminosity range is based on the brightest observed FRBs;
as these are harder to miss, this  end is likely reasonably complete. 

This luminosity basis of our model means wider pulses, on the one hand,
have larger fluence (Eq.~\ref{Eqn: S_peak}) but on the other are
somewhat harder to detect over the noise (Eq.~\ref{eqn:sn}).
Our treatment of these two selection effects allows us to reproduce the CHIME detected widths well. 
The higher-frequency surveys we next plan to include in \frbpoppy alongside CHIME/FRB 
(see Sect.~\ref{sec:si})
 suffer less from scattering effects,
and more directly sample the narrower bursts
\citep[as evidenced in e.g.][]{Pastor2024}.

\subsection{Dispersion by the baryonic IGM}
The $\text{DM}_\text{IGM, slope}$ is relatively poorly constrained in the Euclidean
and delayed SFR models, compared to the
pure SFR model. This is expected as the fraction of low-$z$ FRBs (e.g., $z$ < 0.1) in the former models is higher than
in the SFR model. The $\text{DM}_{\text{IGM, slope}}$ appears in the $\text{DM}_{\text{total}}$ as
$\text{DM}_{\text{IGM, slope}} z$. Therefore, to constrain $\text{DM}_{\text{IGM, slope}}$ better, more high redshift
FRBs are needed. Within  the $1\sigma$ uncertainty, the best-fitting values of $\text{DM}_{\text{IGM, slope}}$ are still consistent with the Macquart relation \citep{Macquart2019,James2022a,James2022b}

\begin{figure*}
    \centering
    \includegraphics[width=1.0\linewidth]{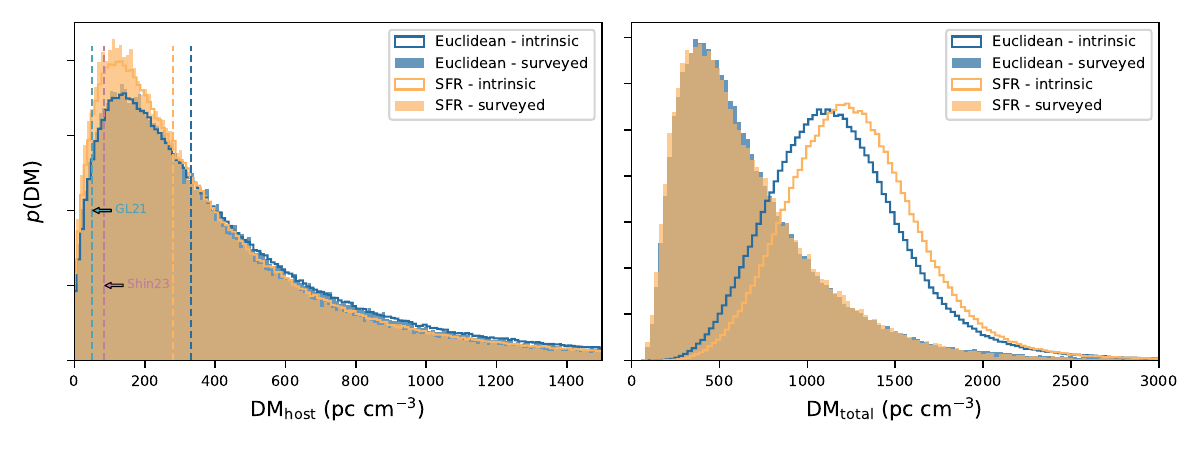}
    \caption{Probability distribution of DM$_{\text{host}}$ (left) in the source frame, and DM$_{\text{total}}$
      (right) before (intrinsic) and after (surveyed) detection in our simulation.
      The median DM$_{\text{host}}$ found
      by \citetalias{Gardenier2021b}, \citet{Shin2023} and this work are indicated with vertical dashed lines. The $y$ axes are in linear scale.}
    \label{Fig: DM}
\end{figure*}

\subsection{Host dispersion measures} 
\label{sec:hostdm}
One of the challenges in understanding the FRB population
is that many of its intrinsic properties are tangled into only a small number of observed distributions.
One such example is the interplay and overlap of the amount of dispersion caused by the host
($\text{DM}_{\text{host}}$), the density of the IGM ($\text{DM}_{\text{IGM,slope}}$), and the distances to the sources.
The first two are mainly constrained 
by the DM$_{\text{total}}$ and the  $w_\text{eff}$ distribution; while the distances are constrained by the
 DM$_{\text{total}}$ and the S/N distribution.
There is an anti-correlation between $\text{DM}_{\text{IGM, slope}}$ and $\text{DM}_{\text{host}}$ in the MCMC contour
plot, most obvious in Fig.~\ref{Fig: MCMC-sfr}.
 
In our simulation, we allowed a $\text{DM}_{\text{host}}$ (including the source contribution) to range over a log-normal
distribution. That distribution is, at the moment, independent of the redshift. Contained in it are the
  possible inclination angles, and host galaxy types, averaging over their evolution between $z=1.5$ and now.
Figure~\ref{Fig: DM} and Table~\ref{Tab: best-fitting parameters}
indicate that the best-fitting values of $\text{DM}_{\text{host}}$ vary only slight between the  different
models. 
In Table~\ref{Tab: best-fitting parameters} we list the mean values we find;
our median  $\text{DM}_{\text{host}}$ in the source rest frame is $\sim$250\,pc cm$^{-3}$ for the SFR model
and $\sim$340\,pc cm$^{-3}$ for  Euclidean model respectively. 

Our values are higher than  assumed earlier (including in
\citetalias{Gardenier2021b}, who find a constant value of $50$\,pc cm$^{-3}$ produced an optimal fit).
Our new results are also different from, e.g., \citet{Shin2023},
who find a median $\text{DM}_{\text{host}}$ of $84^{+69}_{-49}$\,pc cm$^{-3}$ with a standard
deviation of $174^{+319}_{-128}$\,pc cm$^{-3}$.
The difference could, given the anti-correlation mentioned above,
be caused by the value $\text{DM}_{\text{IGM, slope}}$ chosen 
in \cite{Shin2023}, where it is not a free parameter.
But we infer its value to be around 850\,pc cm$^{-3}$ (from their Fig.~10),
not that different from our lowest result from 
the Euclidean model of  $\text{DM}_{\text{IGM, slope}} \simeq$ 780\,pc cm$^{-3}$.

Both \citetalias{Gardenier2021b}
and \citet{Shin2023} find 
median  $\text{DM}_{\text{host}}$ values
lower than the median DM of the known radio pulsars in our own Milky Way 
(where large selection effects against high-DM pulsar detection are at play, see e.g.,
\citealt{2010A&A...509A...7V}).
As FRBs need to escape their host galaxy, which is often massive,
it seems surprising it is found to be this low. 
We find  {median}  $\text{DM}_{\text{host}}$  values of  $\sim$300\,pc cm$^{-3}$ though,
larger than the median for pulsars. 
In our results, the total DM is roughly comprised of equal contributions from
$\text{DM}_{\text{host}}$  and   $\mathrm{DM}_{\mathrm{IGM}}$, but higher values of
 DM$_{\text{total}}$ are generally driven by high host contributions, as visible in Fig.~\ref{Fig: best fitting} (bottom
 panel). 
The median DM for the 6 radio magnetars in our Milky Way, however, is 700\,\pccm.
Our median   $\text{DM}_{\text{host}}$ values are starting to approach this value.
In our Milky Way, the line of sight to magnetars cuts through the dense plane.
Galaxies containing FRB-emitting magnetars \citep[see ][]{2020Natur.587...54C,2022ATel15697....1M}, on the other hand,
are generally seen at an angle, with slightly lower DM. 
We conclude our results support a magnetar origin for FRBs. 

The observed {DM}$_{\text{total}}$ distribution is a convolution of multiple components, many of which evolve over
our sampled redshift range. The source DM contributions may vary strongly depending on the local conditions (supernova
remnants, H{\small II} regions). The galaxy inclination angles matter for lower-redshift disk galaxies, and
the distribution is an average over all these angles. Together, these sum to a {DM}$_{\text{host}}$
distribution as shown in Fig.~\ref{Fig: DM}. This distribution can next be compared to the predictions from
galaxy formation models \citep[e.g.,][]{2023MNRAS.523.5006J} and 
cosmological simulations (for example, \citealt{Orr2024}).

From an overall model perspective, the main difference between our work and \cite{Shin2023}
is that the latter authors
find that the $P_\text{obs}(z)$ distribution peaks
at $\sim$0.36. None of the number density models that we consider (Power-law density,
SFR or delayed SFR) produce a peak
at such high redshift. Indeed, the differences in these redshift distributions are strongly correlated with the different
$\text{DM}_{\text{host}}$ between \cite{Shin2023} and our work. If the FRBs are located at higher $z$, a low
$\text{DM}_{\text{host}}$ is required to reproduce the observed DM curve.
We find our values for   $\text{DM}_{\text{host}}$ are reasonable given the knowledge of the known host galaxies \citep[e.g.,][]{2022AJ....163...69B}.
As \frbpoppy is open source, we encourage researchers to reproduce these results for themselves.

Our results indicate the selection effects against high-DM hosts are not very strong. 
This suggests FRBs from the centers of host galaxies are not strongly disfavored compared to those from the outskirts. 
That is interesting because sources such as magnetars or other young stellar remnants may be found closer to the center,
where the dispersion is higher (see, e.g., \citealt{2020A&A...634A...3V} for such a DM model),
while globular cluster sources are found more at or beyond the host edge, where $\text{DM}_{\text{host}}$ is lower.

\subsection{Statistics}
\label{sub:stat}
We use two statistical tools in this study.
We employ the GoF to determine the global optimum parameters within a model,
and we use the  BIC method to choose between models, and determine  which is the best.
When deriving the log-likelihood function for the GoF, we have
assumed the non-correlation of the three distributions. Although this does not strictly need to be the case, we find
that when the minimum of $A^2_{\text{Total}}$ is reached, the $A^2_{\text{S/N}}$, $A^2_{w_\text{eff}}$ and $A^2_{\text{DM}}$ are all close to minimum as is shown in Table~\ref{Tab: statistic}. We hence obtain the best-fitting values for all parameters simultaneously.

One could argue that for comparing between models, the Bayes factor is an even better tool,
when considering the model complexity. Calculating the model evidence is, however, not included\footnote{\url{https://github.com/dfm/emcee/issues/348}} in the MCMC package \texttt{emcee}
that is the main focus of the current study. We may consider  Bayes factor comparison in future work.
The BIC is sufficient for the problem at hand.
We produced 50 surveyed populations for each model using the best-fitting parameters, and calculate BICs from the averaged GoFs to
reduce the randomness. The results are shown in Table~\ref{Tab: statistic}.
The three delayed SFR models have very close BICs and cannot be distinguished from each other.
They show a hint of evidence over the normal SFR model, but strong evidence over the Euclidean model.

\subsection{Comparison with other studies}
\label{sub:compar}

In Table \ref{Tab: best-fitting parameters} we compare the best-fitting values from our current work against those
reported in  \citetalias{Gardenier2021b}.
In that study, \frbpoppy could not yet produce formal errors; below, we estimate these from the width of
the distributions plotted in Fig.~3 of \citetalias{Gardenier2021b}.
Our value for $\alpha$, $-1.58_{-0.27}^{+0.09}$ falls within the Cycle 1 range of the \citetalias{Gardenier2021b} MC
simulation, but in the following runs, their value moved away, to $-2.2$.
Possibly our full MCMC approach is better able to find the correct global best fit.
Our values for the power-law index are consistent with their near-flat $li$ distribution.

Several reasons may account for the differences.
In the current work, the parameters si, $L_\text{min}$ and $L_\text{max}$ are fixed instead of free.
Furthermore, in \citetalias{Gardenier2021b}, the observed sample size is smaller, and not uniform in its selection
effects.
Next, the A-D test we now use for the GoF emphasizes other difference between distributions than the K-S test did.
Finally, for finding the global maximum,
the current MCMC implementation is a fundamental improvement over 
the \citetalias{Gardenier2021b} MC simulation that ran multiple cycles over a number of parameter subsets. 

\citet{James2022a,James2022b} studied the $z$-DM distribution using the ASKAP and Parkes samples
(with their code \texttt{zDM}\footnote{\url{https://github.com/FRBs/zdm}}),
by  fitting over a DM-redshift distribution model, a power-law energy
distribution and a number density that evolves with modified SFR. They find an estimated maximum FRB energy
of $\log_{10} (E_\text{max}/\text{erg}) = 41.70^{+0.53}_{-0.06}$ (assuming a 1\,GHz bandwidth),
a cumulative luminosity index of
$-1.09^{+0.14}_{-0.10}$ (discussed previously),
an SFR scaling parameter $1.67^{+0.25}_{-0.40}$ and a log-normal distribution of $\text{DM}_\text{host}$ with mean $129^{+66}_{-48}$\,pc\,cm$^{-3}$. 

\cite{Shin2023} build on \citet{James2022b}, but now for CHIME/FRB Catalog 1.
They performed a seven parameter MCMC inference based on the DM$-$Fluence distribution,
in which the likelihood
function was evaluated by comparing the counts per bin between the observed and modeled FRB samples.
We contrasted their findings to ours in Sect.~\ref{sec:hostdm}, specifically on the DM,
the underlying
 number density input model, and the derived redshift distribution.
We note that unlike \citet{James2022b},
 \citet{Shin2023} appear to not forward model the intrinsic pulse widths, or the subsequent scatter broadening.
The authors use selection-corrected fiducial distributions from the injection system instead.
This pulse width has a large, correlated influence on both the source and the telescope behavior.
It strongly affects the telescope S/N, which is the most important metric for detection;
and it has large influence on the emitted burst energy, assuming a given luminosity.  
Given these unmodeled correlations, 
  \citet{Shin2023} necessarily have to limit their sample to avoid parts of the distribution
 where these selection effects occur. In the end, their comparison is based on 225 bursts.
 We model more selection effects, incorporate the entire catalog,
and fit for more observables: S/N, $w_\text{eff}$ and DM$_{\text{total}}$.

\cite{Cui2022} selected a sample of 125 apparently non-repeating FRBs and report their luminosities follow a log-normal distribution.
We, however, find a power-law luminosity function is a good fit.
Possibly the selection criteria used in \cite{Cui2022}, S/N $>12$, introduces a bias against low luminosity FRBs and consequently add weight to the log-normal model.

\cite{Bhattacharyya2022} performed a three-parameter MCMC simulation with an 82 one-off burst sample to study the FRB
spectral index, luminosity spectral index and mean energy.
and \cite{Bhattacharyya2023} next used the CHIME/FRB Catalog 1 to perform a similar study.
We have discussed in Sect.~\ref{sec:si} that a survey at a single frequency
is incapable of pinning down the spectral index,
and allowing it to vary anyway introduces strong covariances in the luminosity function.
This is prominent in the results reported in \cite{Bhattacharyya2023}.
We did not consider a Schechter luminosity function in our MCMC,
and thus cannot compare against the cutoff energy and exponent found in \cite{Bhattacharyya2022}. 
As the fitting of the energy distribution in that work does not appear to account for the selection function of
the telescope, a meaningful comparison would have been difficult, regardless.
 
\cite{Chawla2022} studied the dispersion and scattering properties of the FRBs in
Catalog 1, through a cosmological population synthesis.
They find a DM distribution similar to ours,
  but their scattering time distribution that extends further into the high end (>10\,ms) than our results.
  Still, our pulse width distribution  reproduces  the widths seen by CHIME without the need of such a high-end tail.
   In the left-hand subplot of Fig.~\ref{Fig: pulse width} this is visible from the agreement
between the gray histogram and the green line.
In that figure,
the detected simulated pulses (orange)
are clearly be part of the large-width tail of the intrinsic distribution
(blue).
Scattering, next, is a large further addition to the observed pulse width, as visible in the right sub-panel of 
Fig.~\ref{Fig: pulse width}.

Overall, we find that
our results are based on matching more observables than other work
and we consider them more robust.
The validity of our best models is evidenced by the smaller error budgets we attain.
For example, on the luminosity index
we produce errors of {$-$0.18}, {+0.10}, significantly better than 
the {$-$0.4},  {+0.7} of \citet{Shin2023} or the 
{$-$0.4}, {+0.3} of  \citet{Luo2020}.

\subsection{Snapshot data and result reproducibility}
\label{Sec:data}
 We provide a number of ways for interested astronomers to use our results. We discuss these below.

\subsubsection{Model outcome packages}
\label{sec:data_avail}

To get started immediately,
one can navigate to the
\frbpoppy quickstart section on
Zenodo\footnote{\scriptsize\url{https://doi.org/10.5281/zenodo.11091637}}
 and download \texttt{1\_Day\_FRB\_Sky\_on\_Earth.txt}.
That human- and machine-readable file
contains a simulated 1-day catalog of one-off FRBs, 
that allows users to access the FRB population without installing the entire {\frbpoppy} package.
Out of the $10^{8.3}$ FRBs that happen every day between us and $z=1$ (Sect.~\ref{Sec:nFRBs}),
$3.5 \times 10^{6}$ 
are brighter than 0.01\,Jy\,ms, the best limit in one-off FRB detection
currently \citep{2021ApJ...909L...8N} and in the foreseeable future.
The simulated catalog is produced by the
{\texttt perfect telescope} in \frbpoppy,
free of selection effects,
that observed 4$\uppi$ of sky for 24\,hrs, with minimum detectable fluence 0.01\,Jy\,ms, for the  best-fit no-delay SFR model.

The catalog lists the coordinates, redshift, distance, width, DM, luminosity, fluence,
and other properties for this 1-day snapshot of $3.5 \times 10^{6}$ simulated detected one-off FRBs.
Prospective users can integrate the accompanying python reader into their codes if they wish.

\subsubsection{Open source code}
\label{sec:code_avail}
Version 2.2.0 of {\frbpoppy} is now available through
GitHub\footnote{\scriptsize\url{https://github.com/TRASAL/frbpoppy}}.
% $^*$ \newline$^*$ To be made public at paper acceptance}.
The code continues to support all previous telescopes and population models (including repeating FRBs). 
This latest version includes the  code optimizations and MCMC functionality described in Sect.~\ref{Sec: methods}.

\subsection{Future work}

Since \frbpoppy now takes account into the selection effects important for CHIME/FRB,
logical future steps include inferring the best-fitting populations from the
next CHIME/FRB Catalog release. As   \frbpoppy is open source,
any interested astronomer can take this on. 

Furthermore, a more in-depth and direct comparison
of the various population synthesis codes used in the community
(mentioned in the previous Section)
would be of great interest
if we want to understand code difference, and hence come to a more complete
understanding of the actual population.
Both \frbpoppy and \texttt{zDM} \citep{j_xavier_prochaska_2023_8192369} are open source,
and we urge others to publish reproducibly, too.

Once available, such codes could be 
validated on the same data set as ground truth, with the same metrics.
A collaboration that includes one of us started a similar benchmark for
public FRB {search} codes (see 
\citealt{frbolympics, andre_melo_2020_3903257})
with the goal of improving code and
furthering open science.
A similar comparison of population study codes
would be beneficial for validation, improvement and
sustainability of the projects involved.

The steady increase in the fractional bandwidth employed in FRB surveys
may  begin to warrant including in-band evolution of the
FRB brightness and scatter broadening,
as well as of the telescope sensitivity and beam pattern.
For the current study we find those effects to be small enough,
compared to other factors,
to stay with the quasi-monochromatic approach.
For $t_{\mathrm{scat}}$ and $t_{\mathrm{DM}}$, for example, one could use 
weighted averages over the band. For our 
{median}  $\text{DM}_{\text{total}}$  value of  $\sim$550\,pc cm$^{-3}$
(e.g., Fig.~\ref{Fig: best fitting})
the weighted average over  $t_{\mathrm{scat}}$ is 0.3\,ms 
larger than the central value we use now
(see Eq.~13 in  \citealt{Gardenier2019}).
This is below the CHIME sampling time,
and would not meaningfully influence the surveyed
$t_{\mathrm{scat}}$ distribution (a 0.2$\sigma$ shift in Fig.~\ref{Fig: pulse width}).
The effect on  $t_{\mathrm{DM}}$ is another factor of two smaller.
It would be an improvement to include this in the future none the less, especially for 
searches with even lower frequency telescopes such as LOFAR and Square Kilometre Array (SKA)-Low \citep[e.g.,][]{2014A&A...570A..60C}.

Natural progression of the current results (as mentioned before) could entail
inclusion of Schechter, broken power-law, or log-normal distributions for the luminosity function,
and of the Bayes factor for comparing models.
Modeling of CHIME/FRB repeater FRBs,
and the addition of the other new surveys with tens of FRB detections,
would also be highly interesting.

\section{Conclusions}
\label{Sec: conclusion}
The updated \frbpoppy is able to perform an MCMC on
the entire  CHIME/FRB Catalog 1
and find a single best model for the
birth and evolution of the real, underlying FRB population. 
Where other studies had to curate the input FRBs,
our treatment of selection effects allows us to use all one-off FRBs,
producing better fits, with smaller errors, than before. 

Every Earth day $\sim$$2 \times 10^{8}$ FRBs go off  between us and $z=1$,
and
we make available the mock catalog of the 
$3.5 \times 10^{6}$ that are potentially detectable.
Some of these are nearby and very bright,
and we have presented evidence that a number of these were missed by CHIME/FRB. 

We show that contributions to the DM by the host of several 100s of \pccm
best describe the observed dispersion distribution. We
thus conclude selection effects against relatively high-DM hosts are not very strong. 

We find strong evidence that
the FRB population evolves with look-back time, following star formation,
with a slight preference for a small delay.
We conclude FRBs are produced by the remnants of short-lived stars.
The results from our combination of the largest, single-survey data set
and the most comprehensive population synthesis code  available at the moment
 thus lend population-based  support to the hypothesis that  FRBs are created by neutron stars.

~\\
\begin{spacing}{0.8}
\noindent{\tiny{\it Data availability.}
The code and models used here are available at \url{https://github.com/TRASAL/frbpoppy}.
A snapshot 1-day catalog is available at \url{https://doi.org/10.5281/zenodo.11091637}.
}\end{spacing}%\vspace{-.1ex} 

\begin{acknowledgements}
We thank David Gardenier, Leon Oostrum, and Dany Vohl
for interesting discussions and for their support,
and Pikky Atri, \mbox{Kaustubh} \mbox{Rajwade},  Ralph Wijers and the anonymous referee for suggestions
that improved this work. 
 
This research was supported by the following projects, all financed by the Dutch Research Council (NWO):
Vici research project ``ARGO'' (grant number 639.043.815);
CORTEX (NWA.1160.18.316), under the research programme NWA-ORC;
plus EINF-3624 and EINF-7739
of the research programme ``Computing Time on National Computing Facilities'' hosted by SURF.
\end{acknowledgements}

\bibliographystyle{yahapj}
\bibliography{export-bibtex}

\clearpage
\onecolumn

\begin{appendix}
\section{Summary of parameters and acronyms}
\begin{longtable}{ll}
\caption{The long form and meaning of the parameters and abbreviations used in this work.}\label{Tab:acros}\\
\hline
\multicolumn{1}{l|}{Name}                 & Meaning \\ \hline
\multicolumn{2}{c}{Parameters}                    \\ \hline
\multicolumn{1}{l|}{$A^2$}     & Statistic of Anderson-Darling test  (Eq.~\ref{eq:LA2})   \\
\multicolumn{1}{l|}{$\alpha$}     & Parameter defined by $\alpha \equiv -3/(2 B)$ (Eq.~\ref{eq:alpha})  \\
\multicolumn{1}{l|}{B}      & Power-law index that   modifies the uniform sampling $U(0,1)$ in number density model (Eq.~\ref{eq:Vco}) \\
\multicolumn{1}{l|}{$\beta$}         & Telescope sensitivity degradation factor  (Eq.~\ref{eqn:sn}) \\
\multicolumn{1}{l|}{$D(z)$}         & Proper distance   (Eq.~\ref{eqn:dL})  \\
\multicolumn{1}{l|}{$d_L(z)$}         & Luminosity distance    (Eq.~\ref{eqn:dL})  \\
\multicolumn{1}{l|}{$\text{DM}_{\text{host}}$} & DM contributed by the host galaxy   (Eq.~\ref{eq:DMs})  \\
\multicolumn{1}{l|}{$\text{DM}_{\text{host, mean}}, \text{DM}_{\text{host, std}}$} & mean and standard deviation of host galaxy DM in the log-normal distribution model   \\
\multicolumn{1}{l|}{$\text{DM}_{\text{MW}}$} & DM contributed by the Milky Way  (Eq.~\ref{eq:DMs})   \\
\multicolumn{1}{l|}{$\text{DM}_{\text{IGM}}$} & DM contributed by the intergalactic medium    (Eq.~\ref{eq:DMs})  \\
\multicolumn{1}{l|}{$\text{DM}_{\text{IGM, slope}}$} & Slope of intergalactic medium DM in Macquart relation  \\
\multicolumn{1}{l|}{$\text{DM}_{\text{src}}$} & DM contributed by the circumburst environment    \\
\multicolumn{1}{l|}{$\text{DM}_{\text{total}}$} & Total DM   (Eq.~\ref{eq:DMs})   \\
\multicolumn{1}{l|}{$f_b$}         & Beaming fraction  (Sect.~\ref{Sec:nProgn})  \\
\multicolumn{1}{l|}{$f_d$}         & Delayed formation fraction (Sect.~\ref{sec:res:sfr})   \\
\multicolumn{1}{l|}{$G$}         & Telescope gain   (Eq.~\ref{eqn:sn})   \\
\multicolumn{1}{l|}{$\gamma$}         & Spectral index   (Sect.~\ref{sec:si})   \\
\multicolumn{1}{l|}{$I$}         & Beam intensity   (Eq.~\ref{eqn:sn}) \\
\multicolumn{1}{l|}{$li$}         & Luminosity index in the power-law distribution model  (Sect.~\ref{sec:lum})   \\
\multicolumn{1}{l|}{$L_\mathrm{min}$, $L_\mathrm{max}$}         & Minimum and maximum luminosity in the power-law distribution model (Sect.~\ref{sec:lum})  \\
\multicolumn{1}{l|}{$L_\mathrm{bol}$}         & Bolometric luminosity   (Sect.~\ref{sec:lum})  \\
\multicolumn{1}{l|}{$\ln L$}         & Log likelihood function  (Eq.~\ref{eq:LA2})   \\
\multicolumn{1}{l|}{$\nu_1, \nu_2$}         &   Minimum and maximum observing frequency of a telescope   (Eq.~\ref{Eqn: S_peak})  \\
\multicolumn{1}{l|}{$\nu_\mathrm{low}, \nu_\mathrm{high}$}         & Minimum and maximum frequency range that are used to calculate $\bar{S}_\mathrm{peak}$   (Eq.~\ref{Eqn: S_peak})  \\
\multicolumn{1}{l|}{$p_{\mathrm{K-S}}$}         & $p$-value of Kolmogorov-Smirnov test    \\
\multicolumn{1}{l|}{$\rho$}         & Local volumetric rate of FRBs  (Sect.~\ref{Sec:nProgn})  \\
\multicolumn{1}{l|}{$\bar{S}_\mathrm{peak}$}         & Peak flux density       (Eq.~\ref{Eqn: S_peak})  \\
\multicolumn{1}{l|}{$\tau_d$}         & Formation delay (Sect.~\ref{sec:res:sfr})   \\
\multicolumn{1}{l|}{$t_\mathrm{DM}$}         & Intrachannel dispersion smearing time   (Eq.~\ref{eq:weff})  \\
\multicolumn{1}{l|}{$t_\mathrm{scat}$}         & Scattering time  (Eq.~\ref{eq:weff})   \\
\multicolumn{1}{l|}{$t_\mathrm{samp}$}         & Sampling time  (Eq.~\ref{eq:weff})   \\
\multicolumn{1}{l|}{$T_\mathrm{sys}$}         & System temperature  (Eq.~\ref{eqn:sn})   \\
\multicolumn{1}{l|}{$V_\mathrm{co, FRB}$}         & Comoving volume confined by the redshift of an FRB   (Eq.~\ref{eq:Vco})  \\
\multicolumn{1}{l|}{$V_\mathrm{co, max}$}         & Maximum comoving volume corresponding to a given maximum redshift  (Eq.~\ref{eq:Vco})   \\
\multicolumn{1}{l|}{$w_\mathrm{int}$}         & Intrinsic pulse width   (Eq.~\ref{eq:wint})   \\
\multicolumn{1}{l|}{$w_\mathrm{int, mean}, w_\mathrm{int, std}$}        &  mean and standard deviation of intrinsic pulse width in the log-normal distribution model  (Sect.~\ref{sec:wint})    \\
\multicolumn{1}{l|}{$w_\mathrm{arr}$}         & Pulse width arriving at Earth   (Eq.~\ref{eq:warr})  \\
\multicolumn{1}{l|}{$w_\mathrm{eff}$}         & Effective pulse width    (Eq.~\ref{eq:weff})   \\
\hline
\multicolumn{2}{c}{Acronyms}     \\ 
\hline
\multicolumn{1}{l|}{BIC}         & Bayesian information criterion     \\
\multicolumn{1}{l|}{CDF}         & Cumulative distribution function 
 \\
\multicolumn{1}{l|}{DTD}         & Delay time distribution    \\
\multicolumn{1}{l|}{GoF}         & Goodness-of-fit     \\
\multicolumn{1}{l|}{IGM}         & intergalactic medium    \\
\multicolumn{1}{l|}{ISM}         & Interstellar medium    \\
\multicolumn{1}{l|}{MC}          & Monte Carlo    \\
\multicolumn{1}{l|}{MCMC}        & Markov chain Monte Carlo    \\
\multicolumn{1}{l|}{MLE}         & Maximum likelihood estimation   \\
\multicolumn{1}{l|}{PDF}         & Probability density function    \\
\multicolumn{1}{l|}{RFI}         & Radio frequency interference    \\
\multicolumn{1}{l|}{SEFD}        & System equivalent flux density  \\
\multicolumn{1}{l|}{SFR}         & Star formation rate   \\
\multicolumn{1}{l|}{S/N}         & Signal-to-noise ratio    \\
\multicolumn{1}{l|}{TNS}         & Transient Name Server    \\
\hline
\end{longtable}
\end{appendix}

\end{document}